%% file: main.tex
\documentclass[conference]{IEEEtran}
\IEEEoverridecommandlockouts
\usepackage{cite}
\usepackage{amsmath,amssymb,amsfonts,amsthm}
\usepackage[nolist]{acronym}
\usepackage{algorithmic}
\usepackage{hyperref}
\usepackage{scalerel}
\usepackage{graphicx}
\usepackage{textcomp}
\usepackage[table]{xcolor}
\usepackage{tikz}
\usepackage{nicefrac}
\usepackage{circuitikz}

\usepackage{booktabs}

\usepackage{multirow}

\usepackage{siunitx}

\usepackage{makecell}

\usepackage{xurl}

\newdimen\bpt
\def\mobile#1{\leavevmode 
   \bpt=#1bp \hbox to7\bpt{\kern1\bpt \lower1\bpt\vbox to12\bpt{}%
      \pdfliteral{q #1 0 0 #1 0 0 cm 1 j 2 w 0 0 5 10 re B 
         1 g 1 G  1 w .3 1.8 4.4 7 re B 
         1.5 w 2.5 .2 0 .1 re B .3 w 1.7 10 1.6 0 re B Q}%
      \hss}}

\usepackage{adjustbox}
\usetikzlibrary{calc}
\usetikzlibrary{positioning}
\usepackage{pgfplots}
\usepgfplotslibrary{groupplots}
\usepgfplotslibrary{colorbrewer}
\pgfplotsset{compat = 1.15}

\makeatletter
\let\MYcaption\@makecaption
\makeatother

\usepackage[font=footnotesize]{subcaption}

\makeatletter
\let\@makecaption\MYcaption
\makeatother

\DeclareMathOperator*{\argmax}{arg\,max}

\DeclareMathOperator*{\vectorize}{\mathrm{vec}\,}

\usetikzlibrary{svg.path}

\def\BibTeX{{\rm B\kern-.05em{\sc i\kern-.025em b}\kern-.08em
    T\kern-.1667em\lower.7ex\hbox{E}\kern-.125emX}}

\definecolor{orcidlogocol}{HTML}{A6CE39}
\tikzset{
  orcidlogo/.pic={
    \fill[orcidlogocol] svg{M256,128c0,70.7-57.3,128-128,128C57.3,256,0,198.7,0,128C0,57.3,57.3,0,128,0C198.7,0,256,57.3,256,128z};
    \fill[white] svg{M86.3,186.2H70.9V79.1h15.4v48.4V186.2z}
                 svg{M108.9,79.1h41.6c39.6,0,57,28.3,57,53.6c0,27.5-21.5,53.6-56.8,53.6h-41.8V79.1z M124.3,172.4h24.5c34.9,0,42.9-26.5,42.9-39.7c0-21.5-13.7-39.7-43.7-39.7h-23.7V172.4z}
                 svg{M88.7,56.8c0,5.5-4.5,10.1-10.1,10.1c-5.6,0-10.1-4.6-10.1-10.1c0-5.6,4.5-10.1,10.1-10.1C84.2,46.7,88.7,51.3,88.7,56.8z};
  }
}

\newcommand\orcidicon[1]{\href{https://orcid.org/#1}{\mbox{\scalerel*{
\begin{tikzpicture}[yscale=-1,transform shape]
\pic{orcidlogo};
\end{tikzpicture}
}{|}}}}

\newcommand{\customscriptsize}{\fontsize{7.605pt}{8pt}\selectfont}

\input{acronyms.tex}
\input{corporate_colors.tex}

\begin{document}

\title{CSI Obfuscation: Single-Antenna Transmitters Can Not Hide from Adversarial Multi-Antenna Radio Localization Systems\\
\thanks{This work is supported by the German Federal Ministry of Research, Technology and Space (BMFTR) within the projects Open6GHub (grant no. 16KISK019) and KOMSENS-6G (grant no. 16KISK113).}
}

\author{\IEEEauthorblockN{Phillip Stephan\textsuperscript{\orcidicon{0009-0007-4036-668X}}, Florian Euchner\textsuperscript{\orcidicon{0000-0002-8090-1188}}, Stephan ten Brink\textsuperscript{\orcidicon{0000-0003-1502-2571}} \\}

\IEEEauthorblockA{
Institute of Telecommunications, Pfaffenwaldring 47, University of  Stuttgart, 70569 Stuttgart, Germany \\ \{stephan,euchner,tenbrink\}@inue.uni-stuttgart.de
}
}

\maketitle

\begin{abstract}
The ability of modern telecommunication systems to locate users and objects in the radio environment raises justified privacy concerns.
To prevent unauthorized localization, single-antenna transmitters can obfuscate the signal by convolving it with a randomized sequence prior to transmission, which alters the \ac{CSI} estimated at the receiver.
However, this strategy is only effective against \ac{CSI}-based localization systems deploying single-antenna receivers.
Inspired by the concept of blind multichannel identification, we propose a simple \ac{CSI} recovery method for multi-antenna receivers to extract channel features that ensure reliable user localization regardless of the transmitted signal.
We comparatively evaluate the impact of signal obfuscation and the proposed recovery method on the localization performance of \ac{CSI} fingerprinting, channel charting, and classical triangulation using real-world channel measurements.
This work aims to demonstrate the necessity for further efforts to protect the location privacy of users from adversarial radio-based localization systems.
\end{abstract}


\section{Introduction}\label{sec:intro}
The localization of wireless devices has become indispensable for modern society.
Widely used \acp{GNSS} offer high-accuracy localization, but often fail in complex urban topographies ("street canyons") and indoor environments.
To cope with that issue, the use of wireless communication signals for off-device localization has been investigated \cite{wen_5g_localization_survey}.
Classical, model-based localization methods exploit \ac{ToA} and \ac{AoA} information at the \ac{BS} to estimate user positions.
However, these methods are also susceptible to failure in more complex \ac{NLoS} scenarios.
With the rise of \acp{DNN}, model-agnostic localization methods such as \ac{CSI} fingerprinting have emerged and shown to yield excellent localization performance even in complex environments \cite{savic2015fingerprinting, vieira2017deep, cc_features_ferrand, Arnold2018OnDL}.
A principal drawback of \ac{CSI} fingerprinting is the need for large datasets labeled with expensively acquired ground truth positions for training.
As a self-supervised alternative, channel charting was proposed in \cite{studer_cc}, which aims to learn a physically meaningful, low-dimensional representation of the radio environment by exploiting inherent neighborhood relationships within the measured \ac{CSI}.

\begin{figure}
    \centering
    \input{fig/adversarial_localization_concept}
    \vspace{-0.2cm}
    \caption{Alice, a single-antenna \ac{UE}, transmits a signal $\mathbf{\tilde s}$ (obfuscated by sequence $\mathbf{\tilde v}$) to Bob, a multi-antenna \ac{BS}. Eve, a proximate adversarial multi-antenna \ac{BS}, can intercept this signal and thus locate Alice without her consent.}
    \label{fig:adversarial_localization_concept}
    \vspace{-0.5cm}
\end{figure}
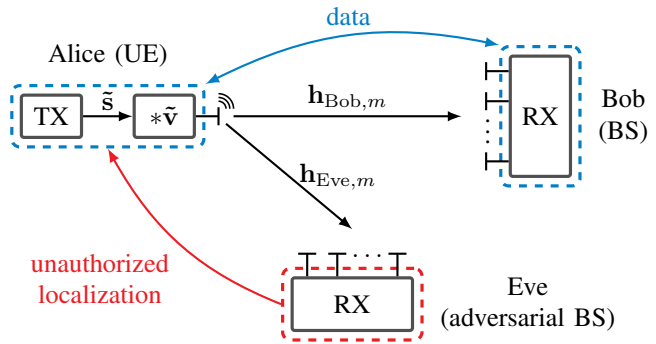

Regardless of the method, localization services generally raise justified privacy concerns \cite{wicker_location_privacy}.
While state-of-the-art cellular devices offer users control over application permissions for location data provided by \acp{GNSS} \cite{orman_mobile_privacy}, they are at mercy of the good intentions of proximate wireless receivers, since wireless communication standards allow practically anyone to gather the \ac{UE}'s \ac{CSI} \cite{gringoli_free_your_csi}.
The disclosure of unencrypted information such as \ac{MAC} addresses facilitates the identification and tracking of \acp{UE} by adversarial sniffers.
Since higher-layer strategies such as \ac{MAC} address randomization have proven insufficient to cope with this vulnerability \cite{martin_mac_randomization, et_tu_alexa}, additional investigations are made on physical layer methods for \acp{UE} to prevent being located by unauthorized entities.
Explicitly addressing \ac{DNN}-based localization systems, previous work has proposed to obfuscate the frequency-domain \ac{OFDM} symbols by multiplying them with a randomized signal \cite{leave_no_trace_wifi_obfuscation, cominelli_csi_randomization, cominelli_csi_obfuscation}.
Highlighting the potential impact on quality of service, the authors of \cite{studer_csi_obfuscation} proposed convolving time-domain signals with random, finite-length obfuscation sequences.
They conjectured that multi-antenna receivers are generally harder to attack since all antenna elements observe the same obfuscation signal.
The literature indeed provides several blind multichannel identification methods \cite{xu_blind_multichannel_estimation,moulines_blind_multichannel_estimation} aiming to reconstruct the \ac{CSI} from signals received at multi-antenna \acp{BS} without pilot symbols.
However, these methods often require knowledge of the channel order and get particularly complex for a large number of \ac{BS} antennas.

\begin{table}
    \caption{Symbols and notations used in this paper}\label{tab:notations}
    \vspace{-0.18cm}
    \centering
    \begin{tabular}{r | l}
        $\mathbf A$, $\mathbf b$ & \begin{tabular}{@{}l@{}}Bold letters: Uppercase for matrices and tensors (here \vspace{-0.07cm}\\ as multidimensional arrays), lowercase for vectors\end{tabular}\\
        $m, N$ & Italic uppercase or lowercase letters: Scalars\\
        $\mathbf{A}^{(l)}$ & Superscript letters: indexing time instant $l$ of tensor $\mathbf{A}$\\
        $\mathbf{A}_{ijk}$ & \begin{tabular}{@{}l@{}}Subscript letters: indexing elements \vspace{-0.07cm}\\ along axes $i,j,k$ of tensor $\mathbf{A}$\end{tabular}\\
        \begin{tabular}{l@{}l@{}l@{}}$\mathbf A_{i::}$ \vspace{-0.07cm}\\$\mathbf A_{ij:}$ \vspace{-0.07cm}\\$\mathbf A_{i:k}$ \end{tabular} & \begin{tabular}{@{}l@{}l@{}}Sub-matrix (and sub-vector) of elements in $i$\textsuperscript{th} entry \vspace{-0.07cm}\\of the first dim. (and $j$\textsuperscript{th} entry of the second dim. \vspace{-0.07cm}\\or $k$\textsuperscript{th} entry of the third dim.) of tensor $\mathbf A$\end{tabular}\\
        $\lVert \mathbf b \rVert$ & Euclidean norm of vector $\mathbf b$ \vspace{0.02cm}\\
        $\mathbf A^\mathrm{H}$, $\mathbf b^\mathrm{H}$ & Conjugate transpose of matrix $\mathbf A$ (or vector $\mathbf b$)\\
        $\mathbf a \odot \mathbf b$, $\mathbf a \oslash \mathbf b$ & Hadamard product (division) of vectors $\mathbf a$ and $\mathbf b$ \\
        $f * g$ & Convolution of signals $f$ and $g$ \\
    \end{tabular}
    \vspace{-0.5cm}
\end{table}

\subsection{Contributions}
\vspace{-0.1cm}
In contrast to blind multichannel identification, this work proposes a simple \ac{CSI} recovery method that allows multi-antenna \acp{BS} to reconstruct certain channel features from received signals.
It eliminates signal components that are correlated across the individual \ac{BS} antennas, namely the transmitted signal and common channel features such as baseband filters or other hardware effects, which are usually constant over time and thus contain minimal information about the \ac{UE}'s location.
Hence, the reconstructed features preserve information about the individual signal propagation paths while being robust against \ac{UE}-side signal obfuscation, making them particularly useful for \ac{DNN}-based \ac{UE} localization.
We investigate the impact of signal obfuscation and the recovery method on the localization performance for \ac{CSI} fingerprinting, channel charting and classical triangulation, and critically discuss its applicability in different scenarios\footnote{The datasets and source code used in this work are publicly available at \href{https://github.com/phillipstephan/Adversarial-Radio-Localization-under-CSI-Obfuscation}{\customscriptsize{github.com/phillipstephan/Adversarial-Radio-Localization-under-CSI-Obfuscation}}}.

\subsection{Outline}
\vspace{-0.1cm}
The remainder of this paper is structured as follows.
Section~\ref{sec:dataset_system_model} details the threat model and the specifications of our dataset.
A short description of the applied \ac{CSI}-based localization methods, namely classical triangulation, \ac{CSI} fingerprinting, and channel charting, is given in Section~\ref{sec:csi_localization}.
Subsequently, the \ac{UE}-side signal obfuscation is described in Section~\ref{sec:csi_obfuscation}, followed by the introduction of our \ac{CSI} recovery method in Section~\ref{sec:csi_recovery}.
The impact of signal obfuscation and the recovery method on the localization performance is evaluated in Section~\ref{sec:results}.
Finally, Section~\ref{sec:conclusion} provides a conclusion and an outlook on possible future research activities.
The symbols and notations used in this paper are shown in Table~\ref{tab:notations}.

\subsection{Limitations}
\vspace{-0.1cm}
The efficacy of our recovery method depends on a sufficiently large number of spatially separated \ac{BS} antennas.
Experiments in Section~\ref{sec:results} show an increasingly negative impact of this method on localization performance with fewer \ac{BS} antennas.
Furthermore, this work only considers single-antenna transmitters, while multi-antenna transmitters are expected to have better capabilities to trick localization systems due to their directivity.
Moreover, we do not claim to recover the actual channel coefficients at the \ac{BS}, but certain features that can be used by malicious \acp{BS} to locate nearby \acp{UE}.
\vspace{-0.1cm}

\section{Threat Model and Dataset}\label{sec:dataset_system_model}

\begin{figure*}
    \centering
    \begin{subfigure}[b]{0.33\textwidth}
        \centering
        \includegraphics[width=0.95\textwidth, trim = 30 200 30 0, clip]{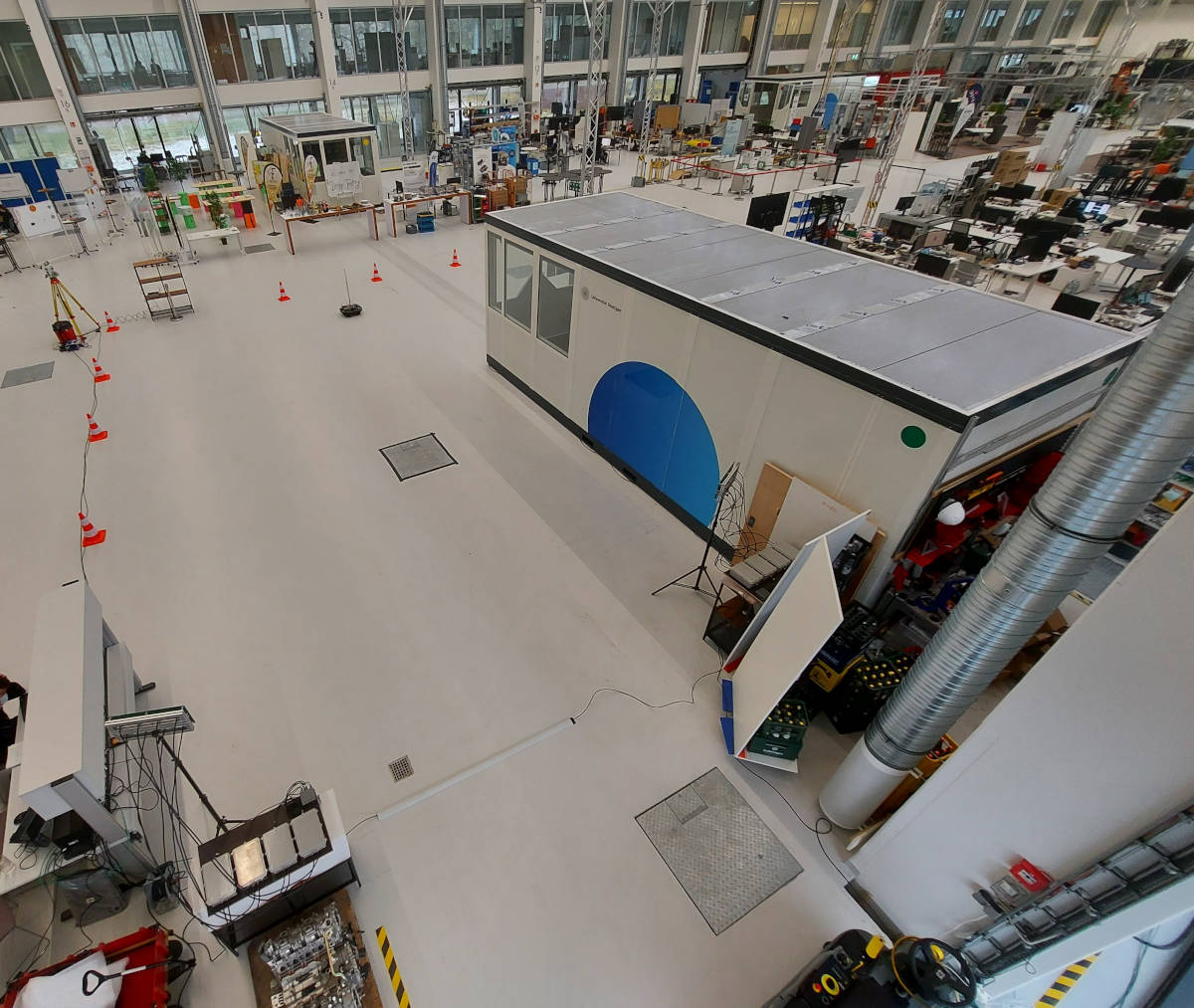}
        \vspace{0.2cm}
        \caption{}
    \end{subfigure}
    \begin{subfigure}[b]{0.32\textwidth}
        \centering
        \begin{tikzpicture}
            \begin{axis}[
                width=0.729\columnwidth,
                height=0.6\columnwidth,
                scale only axis,
                xmin=-15.5,
                xmax=6.1,
                ymin=-18.06,
                ymax=-1.5,
                xlabel = {Coordinate $x_1 ~ [\mathrm{m}]$},
                ylabel = {Coordinate $x_2 ~ [\mathrm{m}]$},
                ylabel shift = -8 pt,
                xlabel shift = -4 pt,
                xtick={-10, -6, -2, 2}
            ]
                \addplot[thick,blue] graphics[xmin=-14.5,ymin=-17.06,xmax=4.1,ymax=-1.5] {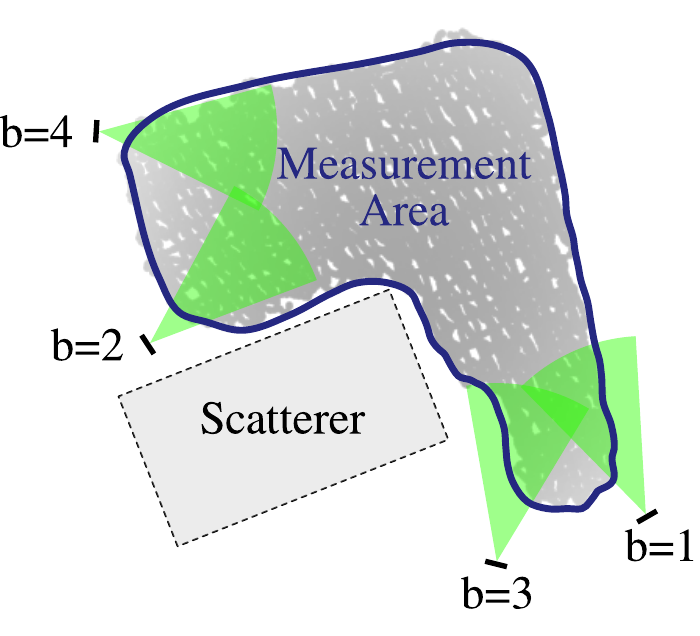};
            \end{axis}
        \end{tikzpicture}
        \vspace{-0.2cm}
        \caption{}
        \label{fig:labelled-area}
    \end{subfigure}
    \begin{subfigure}[b]{0.32\textwidth}
        \centering
        \begin{tikzpicture}
            \begin{axis}[
                width=0.6\columnwidth,
                height=0.6\columnwidth,
                scale only axis,
                xmin=-12.5,
                xmax=2.5,
                ymin=-14.5,
                ymax=-1.5,
                xlabel = {Coordinate $x_1 ~ [\mathrm{m}]$},
                ylabel = {Coordinate $x_2 ~ [\mathrm{m}]$},
                ylabel shift = -8 pt,
                xlabel shift = -4 pt,
                xtick={-10, -6, -2, 2}
            ]
                \addplot[thick,blue] graphics[xmin=-12.5,ymin=-14.5,xmax=2.5,ymax=-1.5] {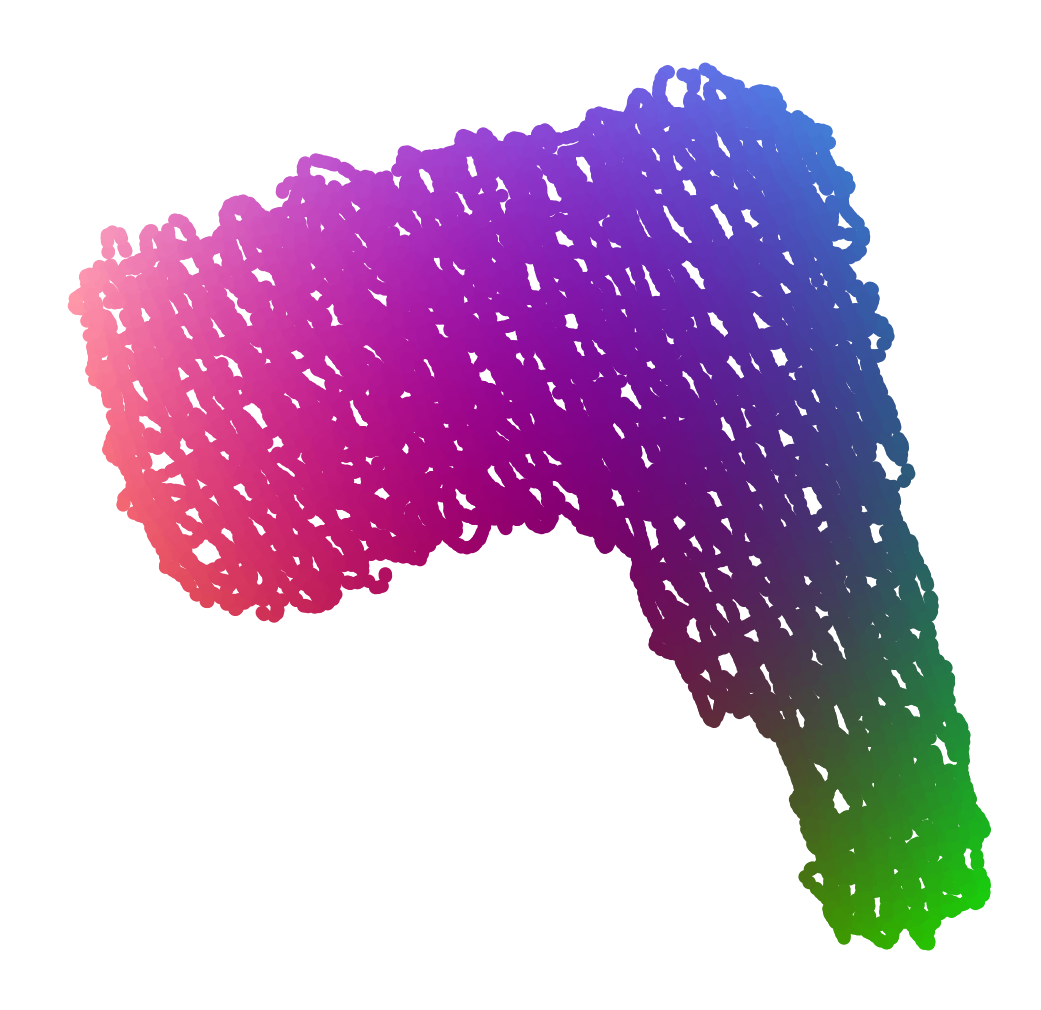};
            \end{axis}
        \end{tikzpicture}
        \vspace{-0.2cm}
        \caption{}
        \label{fig:groundtruth-map}
    \end{subfigure}
    \vspace{-0.3cm}
    \caption{The figure shows (a) a photograph of the measurement environment, (b) a top view map with antenna arrays drawn to scale as black rectangles and their viewing direction indicated by the green sectors, and (c) a scatter plot of arbitrarily colorized ``ground truth'' positions of datapoints in $\mathcal D_\mathrm{train}$.}
    \vspace{-0.2cm}
    \label{fig:industrial_environment}
\end{figure*}

\subsection{Threat Model}\label{sec:threat_model}
We consider the scenario illustrated in Fig.~\ref{fig:adversarial_localization_concept}: Alice, a single-antenna \ac{UE}, transmits a signal $\mathbf{\tilde s}$ to Bob, a nearby multi-antenna \ac{BS}, based on which the individual channel realizations $\mathbf{h}_{\mathrm{Bob},m}$ can be estimated.
Eve, a proximate adversarial multi-antenna \ac{BS}, can sniff the signal and estimate the respective channel realizations $\mathbf{h}_{\mathrm{Eve},m}$, which can be used for unauthorized localization.
To avoid that, Alice obfuscates the signal by convolving it with a random sequence $\mathbf{\tilde v}$ prior to transmission, which effectively alters the estimated channel realizations and thus impairs the localization system.
Making use of the fact that $\mathbf{\tilde v}$ is observed at all receiver antennas, we propose a method for multi-antenna receivers to extract channel features independent of $\mathbf{\tilde v}$ from the received signal, which can be used to still locate Alice (detailed in Section~\ref{sec:csi_recovery}).

\subsection{Dataset}\label{sec:dataset}
We consider the \emph{dichasus-cf0x} dataset \cite{dataset-dichasus-cf0x} measured in an industrial environment with our \ac{DICHASUS} \cite{dichasus2021}.
The system involves a mobile single-antenna \ac{UE} and a distributed massive \ac{MIMO} \ac{BS} with $B=4$ distributed uniform planar arrays, each comprising $M_\mathrm{r} \times M_\mathrm{c} = 2 \times 4$ patch antennas (2 rows, 4 columns), with all \ac{BS} antennas synchronized in frequency, time and phase \cite{geometry_phase_time_sync}.
The dataset contains the frequency-domain channel coefficients $\mathbf{H}^{(l)} \in \mathbb{C}^{B \times M_\mathrm{r} \times M_\mathrm{c} \times N_\mathrm{sub}}$ for all $B\times M_\mathrm{r} \times M_\mathrm{c}$ \ac{BS} antennas and $N_\mathrm{sub} = 64$ \ac{OFDM} subcarriers (subsampled), as well as corresponding ground truth \ac{UE} positions $\mathbf{x}^{(l)} \in \mathbb{R}^2$ and timestamps $t^{(l)} \in \mathbb{R}$ for each time instance $l = 1, \ldots, L$:
\[
    \text{Dataset}: \mathcal D = \left\{ \left(\mathbf H^{(l)}, \mathbf x^{(l)}, t^{(l)} \right) \right\}_{l = 1, \ldots, L}
\]
For training and evaluation, a training set $\mathcal D_\mathrm{train}$ and a test set $\mathcal D_\mathrm{test}$ are distinctively sampled from the same measurement area in \emph{dichasus-cf02}, \emph{dichasus-cf03} and \emph{dichasus-cf04}, with cardinalities $L_\mathrm{train}=20851$ and $L_\mathrm{test}=20851$, respectively.

\section{CSI-based Localization}\label{sec:csi_localization}
\subsection{Baseline: Classical Triangulation}\label{sec:classical_triangulation}
Although the primary focus of this work lies on \ac{DNN}-based localization, we implement a classical triangulation baseline similar to \cite{asilomar2023}, which we do not expect to be significantly affected by \ac{CSI} obfuscation.
At first, the azimuth covariance matrix for each array $b$ and time instant $l$ is computed as
\[
    \mathbf R_b^{(l)} = \sum_{m_\mathrm{r} = 1}^{M_\mathrm{r}} \sum_{n = 1}^{N_\mathrm{sub}} \left(\mathbf H_{bm_\mathrm{r}:n}^{(l)}\right) \left(\mathbf H_{bm_\mathrm{r}:n}^{(l)}\right)^\mathrm{H}
\]
and used by the root-MUSIC algorithm to estimate the respective azimuth \acp{AoA} $\hat{\alpha}_b^{(l)}$.
Assuming the errors of $\hat{\alpha}_b^{(l)}$ adhere to a wrapped normal distribution, approximated by the von Mises distribution, the \ac{AoA} likelihood function is given as
\begin{equation}
    \mathcal L_\mathrm{tri}^{(l)}(\mathbf x) = \prod_{b = 1}^B \frac{\exp \left( \kappa_b \cos \left( \angle_\mathrm{az}(\mathbf x - \mathbf p_b, \mathbf n_b) - \hat \alpha_b^{(l)} \right) \right)}{2 \pi I_0(\kappa_b^{(l)})},
    \label{eq:aoalikelihood}
\end{equation}
where $\angle_\mathrm{az}(\mathbf x - \mathbf p_b, \mathbf n_b)$ is the azimuth angle between the position $\mathbf x$ relative to array $b$ located at $\mathbf p_b$ with respect to its normal vector $\mathbf n_b$.
$I_0$ denotes the modified Bessel function of the first kind of order zero, and $\kappa_b$ is a concentration parameter derived from a heuristic linked to the delay spread observed at array $b$.
Finally, the position estimate $\mathbf { \hat x }^{(l)}$ is obtained through \ac{MLE} as
\[
    \mathbf { \hat x }^{(l)} = \argmax_{\mathbf x} \mathcal L^{(l)}_\mathrm{tri}(\mathbf x).
\]

\subsection{CSI Fingerprinting}\label{sec:fingerprinting}

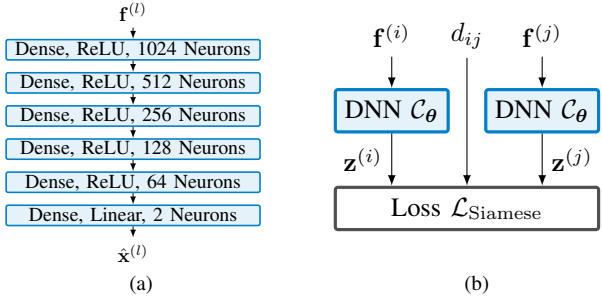
\begin{figure}
    \centering
    \begin{subfigure}[b]{0.49\columnwidth}
        \centering
        \scalebox{0.75}{
        \input{fig/dnn_structure}
        }
        \vspace{-0.15cm}
        \caption{}
        \label{fig:dnn_structure}
    \end{subfigure}
    \begin{subfigure}[b]{0.49\columnwidth}
        \centering
        \scalebox{1.0}{
        \input{fig/siamese_structure}
        }
        \vspace{0.35cm}
        \caption{}
        \label{fig:siamese_structure}
    \end{subfigure}
    \vspace{-0.7cm}
    \caption{Neural network structures: (a) \ac{DNN} for \ac{CSI} fingerprinting or as \ac{FCF}, and (b) Siamese network for the training process of channel charting.}
    \label{fig:neural_networks}
\end{figure}

\ac{CSI} fingerprinting is a simple, yet effective method for indoor localization, which involves training a \ac{DNN} on a large \ac{CSI} dataset labeled with ground truth positions.
The used \ac{DNN}, whose structure is illustrated in Fig.~\ref{fig:dnn_structure}, can be seen as a function $\mathbf {\hat x}^{(l)} = \mathcal G_{\boldsymbol{\theta}}(\mathbf f^{(l)})$ with trainable parameters $\boldsymbol{\theta}$ that maps an input \ac{CSI} feature vector $\mathbf f^{(l)}$ to a position estimate $\mathbf {\hat x}^{(l)}$.
The \ac{CSI} features are derived in a separate feature engineering step from $\mathbf {H}^{(l)}$ and ideally contain its meaningful information while neglecting the redundant.
We compute our features as in \cite{stephan2024angle} based on time-domain \ac{CSI}, which is obtained by computing the \ac{FFT} over the subcarrier axis.
The majority of the signal power is typically observed within $N_\mathrm{tap} = \tau_\mathrm{max} - \tau_\mathrm{min}$ time taps.
Based on the delay spread and bandwidth, we assume $\tau_\mathrm{min} = 27$ and $\tau_\mathrm{max} = 40$.
This subset of taps is extracted from the time-domain \ac{CSI} and stored as $\mathbf{\Tilde H}^{(l)} \in \mathbb C^{B \times M_\mathrm{r} \times M_\mathrm{c} \times N_\mathrm{tap}}$.
To capture angle-delay information, we compute sample autocorrelations across the antenna dimensions for each array $b$ and time tap $\tau$ to get the features $\mathbf F^{(l)}_{bt} = \left( \vectorize \mathbf {\Tilde H}_{b::t}^{(l)} \right) \left( \vectorize \mathbf {\Tilde H}_{b::t}^{(l)} \right)^\mathrm{H} \in \mathbb C^{\left(M_\mathrm{r} \cdot M_\mathrm{c}\right) \times \left(M_\mathrm{r} \cdot M_\mathrm{c}\right)}$.
The final feature vector $\mathbf f^{(l)} \in \mathbb R^{2 \cdot B \cdot \left(M_\mathrm{r} \cdot M_\mathrm{c}\right)^2 \cdot N_\mathrm{tap}}$ is obtained by vectorizing $\mathbf F^{(l)}_{bt}$ and stacking its real and imaginary part.
Supervised training is performed on $\mathcal D_\mathrm{train}$ by applying the \ac{MSE} loss between the position estimates $\mathbf {\hat x}^{(l)}$ and the labels $\mathbf {x}^{(l)}$.
Once trained, the \ac{DNN} can infer position estimates from previously unseen \ac{CSI} samples measured in the same environment, as found in $\mathcal D_\mathrm{test}$.

\subsection{Channel Charting}\label{sec:dissimilarity_cc}

Channel charting \cite{studer_cc} leverages manifold learning to learn a physically meaningful low-dimensional representation of the radio environment by preserving inherent similarity relationships between \ac{CSI} samples.
This paper applies \emph{dissimilarity metric-based channel charting} as in \cite{stephan2024angle}.
At first, we compute dissimilarities ("pseudo-distances") $d_{ij}$ between all pairs of datapoints $i$ and $j$ in $\mathcal D_\mathrm{train}$ based on the \emph{geodesic, fused} dissimilarity metric \cite{stephan2024angle}, which exploits angle-delay profile features from time-domain \ac{CSI} and timestamp differences, followed by a shortest path algorithm to obtain globally meaningful dissimilarities.
Then, the \acf{FCF} $\mathbf {z}^{(l)} = \mathcal C_{\boldsymbol{\theta}}(\mathbf f^{(l)})$ (implemented as a \ac{DNN} with trainable parameters $\boldsymbol{\theta}$) is learned, which maps the \ac{CSI} feature $\mathbf{f}^{(l)}$ to the channel chart representation $\mathbf{z}^{(l)} \in \mathbb{R}^2$.
\ac{CSI} features and \ac{DNN} architecture (Fig.~\ref{fig:dnn_structure}) are adopted from Section~\ref{sec:fingerprinting}.
During training, the \ac{DNN} is embedded in a Siamese network (Fig.~\ref{fig:siamese_structure}), which allows the \ac{DNN} to process two input feature vectors $\mathbf{f}^{(i)}$ and $\mathbf{f}^{(j)}$ concurrently.
The estimated channel chart positions $\mathbf{z}^{(i)}$ and $\mathbf{z}^{(j)}$ are optimized such that their Euclidean point-to-point distance aligns with the respective dissimilarity $d_{ij}$, which is achieved by the Siamese loss
\begin{equation}
\mathcal{L}_\mathrm{Siamese}=\sum\nolimits_{i=1}^{L-1}\sum\nolimits_{j=i+1}^L \frac{\left(d_{ij}-\Vert\mathbf{z}^{(i)}-\mathbf{z}^{(j)}\Vert\right)^2}{d_{ij} + \beta},
\label{eq:siameseloss}
\end{equation}
where $L$ is the number of training samples and the hyperparameter $\beta$ weights the absolute squared error and the normalized squared error.
The channel chart positions $\{\mathbf{z}^{(l)}\}_{l=1}^L$ ideally preserve both local neighborhood relationships and the global structure of the radio environment.
Note that channel chart positions are typically not expressed within a physical coordinate frame, but rather in a transformed version of it.

\section{User-side Attack Model}\label{sec:csi_obfuscation}
To prevent being located, a single-antenna \ac{UE} applies the random attack as in \cite{studer_csi_obfuscation}, where the time-domain signal $\mathbf{\tilde s}^{(l)} \in \mathbb{C}^{N_\mathrm{sub}}$ is convolved with a random obfuscation sequence $\mathbf{\tilde v}^{(l)} \in \mathbb{C}^{L_\mathrm{v}}$ of length $L_\mathrm{v} = 16$ prior to transmission.
The obfuscation sequence is randomly generated as $\mathbf{\tilde v}^{(l)} = a_i^{(l)} e^{j\phi_i^{(l)}}$ with $a_i^{(l)} \sim \mathcal{U}\left(\left[0,1\right]\right)$ and $\phi_i^{(l)} \sim \mathcal{U}\left(\left[0,2\pi\right)\right)$ for $i = 1,\ldots, L_\mathrm{v}$, and normalized as $\lVert \mathbf{\tilde v}^{(l)}\rVert = 1$.
By applying a \ac{DFT} to a zero-padded version of $\mathbf{\tilde v}^{(l)}$, the respective frequency-domain transfer function $\mathbf{v}^{(l)} = \sqrt{N_\mathrm{sub}} \mathbf{F}\left[\mathbf{\tilde v}^{(l)T}, \mathbf{0}_{\left(N_\mathrm{sub}-L_\mathrm{v}\right)}^T\right]^T \in \mathbb{C}^{N_\mathrm{sub}}$ is obtained, where $\mathbf{F} \in \mathbb{C}^{N_\mathrm{sub} \times N_\mathrm{sub}}$ represents the \ac{DFT} matrix.
Given that a convolution in time-domain corresponds to a multiplication in frequency-domain, the \ac{BS} receives the signal at each antenna $m = 1, \ldots, M = \left(B \cdot M_\mathrm{r} \cdot M_\mathrm{c}\right)$ as
\[
    \mathbf{y}_{m}^{(l)} = \left(\mathbf{v}^{(l)} \odot \mathbf{h}_{m}^{(l)}\right) \odot \mathbf{s}^{(l)} + \mathbf{n}_{m}^{(l)} = \mathbf{o}_{m}^{(l)} \odot \mathbf{s}_{m}^{(l)} + \mathbf{n}_{m}^{(l)} \in \mathbb{C}^{N_\mathrm{sub}},
\]
with the frequency-domain signal $\mathbf{s}^{(l)}$, the physical channel between \ac{UE} and $m$-th antenna $\mathbf{h}_{m}^{(l)}$, and the zero-mean white Gaussian noise $\mathbf{n}_{m}^{(l)}$.
Thus, the effective channel at antenna $m$ is observed as $\mathbf{o}_m^{(l)} = \mathbf{v}^{(l)} \odot \mathbf{h}_m^{(l)} \in \mathbb{C}^{N_\mathrm{sub}}$.
A visualization of exemplary time-domain channel realizations and their respective obfuscated version is given in Fig.~\ref{fig:csi_time_domain_RAW} and Fig.~\ref{fig:csi_time_domain_OBFUSCATED}.

\section{CSI Recovery at the Receiver}\label{sec:csi_recovery}

\begin{figure*}
    \begin{subfigure}[b]{0.49\textwidth}
    \begin{adjustbox}{width=\linewidth}
    \input{fig/csi_time_domain_RAW}
    \end{adjustbox}
    \vspace{-0.6cm}
    \caption{Original \ac{CSI}}
    \label{fig:csi_time_domain_RAW}
    \end{subfigure}
    \hspace{0.02\textwidth}
    \begin{subfigure}[b]{0.49\textwidth}
    \begin{adjustbox}{width=\linewidth}
    \input{fig/csi_time_domain_RAW_RECOVERED}
    \end{adjustbox}
    \vspace{-0.6cm}
    \caption{Original \ac{CSI} (recovered)}
    \label{fig:csi_time_domain_RAW_RECOVERED}
    \end{subfigure}
    \begin{subfigure}[b]{0.49\textwidth}
    \begin{adjustbox}{width=\linewidth}
    \input{fig/csi_time_domain_OBFUSCATED}
    \end{adjustbox}
    \vspace{-0.6cm}
    \caption{Obfuscated \ac{CSI}}
    \label{fig:csi_time_domain_OBFUSCATED}
    \end{subfigure}
    \hspace{0.02\textwidth}
    \begin{subfigure}[b]{0.49\textwidth}
    \begin{adjustbox}{width=\linewidth}
    \input{fig/csi_time_domain_RECOVERED}
    \end{adjustbox}
    \vspace{-0.6cm}
    \caption{Obfuscated \ac{CSI} (recovered)}
    \label{fig:csi_time_domain_RECOVERED}
    \end{subfigure}
    \vspace{-0.8cm}
    \caption{Exemplary time-domain channel realizations at an arbitrary time instant: Original, Obfuscated and Recovered.}
    \vspace{-0.2cm}
    \label{fig:csi_time_domain}
\end{figure*}
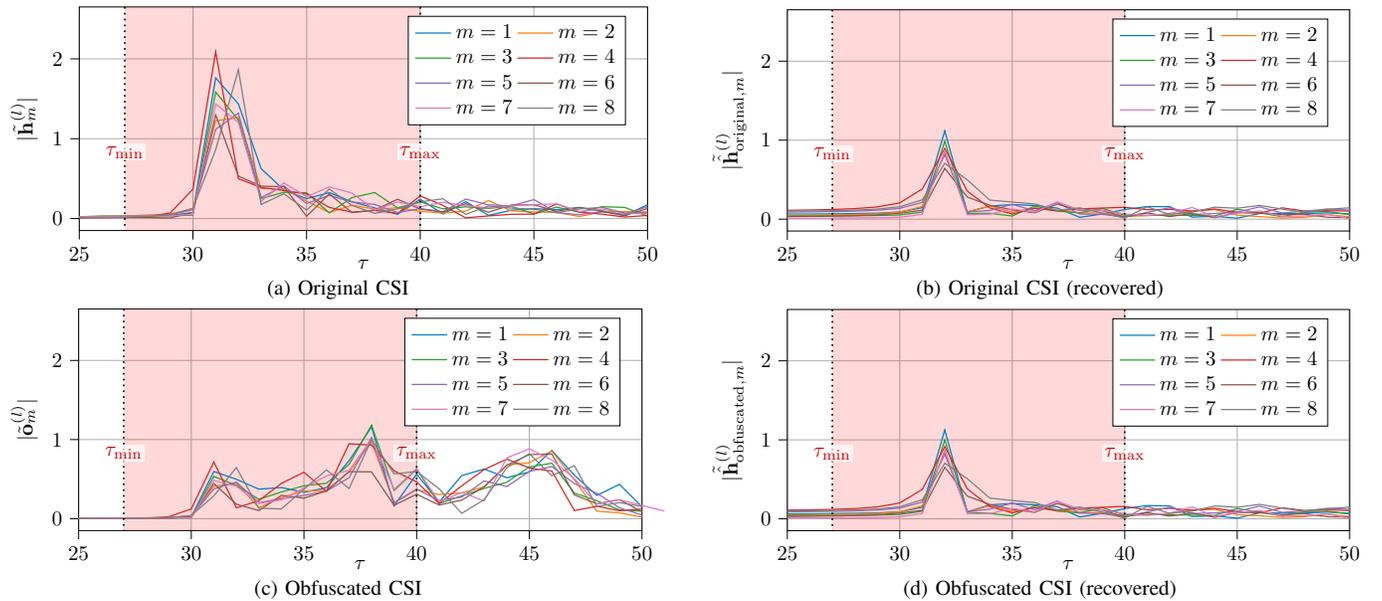

\ac{DNN}-based localization systems are trained to associate \ac{CSI} features with corresponding position estimates.
At a given time instant, \ac{BS} antenna $m$ observes the obfuscated \ac{CSI} as
\[
\mathbf{o}_m = \mathbf{v} \odot \mathbf{h}_m \in \mathbb{C}^{N_\mathrm{sub}}.
\]
Since $\mathbf{v}$ is randomly generated for each datapoint, the \ac{DNN} did almost certainly never encounter similar observations during training and is therefore not capable of estimating any meaningful position.
It is generally important for \ac{DNN}-based localization that the observed \ac{CSI} is unambiguous for any given \ac{UE} position, and simultaneously time-invariant and independent of the signal $\mathbf{v}$.
In the following, we broaden the definition of $\mathbf{v}$ to be any unknown communication signal from the \ac{UE} that does not contain zeros at the observed frequencies.
The objective of our proposed recovery method is to estimate certain channel features $\mathbf{\hat h}_m$ meeting the aforementioned conditions, purely from the observations $\mathbf{o}_m$ (and without knowledge about $\mathbf{v}$ or $\mathbf{h}_m$).
This can be seen as a relaxed form of blind multichannel identification \cite{xu_blind_multichannel_estimation,moulines_blind_multichannel_estimation}, which aims to estimate the exact channel realizations $\mathbf{h}_m$ without pilot symbols.
Given that $\mathbf{v}$ is common to all \ac{BS} antennas, and assuming a sufficiently large number of spatially separated \ac{BS} antennas such that their individual signal propagation paths are mostly independent, the autocorrelation matrix $\mathbf{R} \in \mathbb{C}^{N_\mathrm{sub} \times N_\mathrm{sub}}$ of the observations $\mathbf{o}_m$ can be expressed as
\[
    \mathbf{R} = \mathbb{E}_m \left[\left(\mathbf{v} \odot \mathbf{h}_m\right) \left(\mathbf{v} \odot \mathbf{h}_m\right)^H \right] = \left(\mathbf{v} \mathbf{v}^H\right) \odot \mathbf{R}_{hh},
\]
where $\mathbf{R}_{hh} = \mathbb{E}_m \left[\mathbf{h}_m \mathbf{h}_m^H \right]$ is the autocorrelation matrix of the physical channel realizations $\mathbf{h}_m$.
At first, we aim to find a common spectral pattern $\mathbf{\hat w}$ that is present at all antennas by solving the optimization problem:
\begin{equation}\label{eq:optimization_problem}
    \begin{aligned}
        \mathbf{\hat w} &= \argmax_\mathbf{w} \mathbf{w}^H \mathbf{R} \mathbf{w} \quad \textrm{s.t.} \quad \lVert \mathbf{w}\rVert = 1\\
        &= \argmax_\mathbf{w} \sum_m \mathbf{w}^H \left(\mathbf{o}_m \mathbf{o}_m^H\right) \mathbf{w}\\
        &= \argmax_\mathbf{w} \sum_m \lVert \mathbf{w}^H \mathbf{o}_m\rVert^2.
    \end{aligned}
\end{equation}
Figuratively speaking, we are searching for the vector that maximizes correlation with respect to all observed channel realizations, which is equivalent to finding the principal eigenvector of $\mathbf{R}$.
Let $\mathbf{R}_{hh} = \sum_{n=1}^{N_\mathrm{sub}} \lambda_n \boldsymbol{\vartheta}_n \boldsymbol{\vartheta}_n^H$ be the eigendecomposition of $\mathbf{R}_{hh}$, where $\lambda_n$ is the $n$-th eigenvalue and $\boldsymbol{\vartheta}_n$ the $n$-th eigenvector.
Consequently,
\begin{equation}
    \begin{aligned}
        \mathbf{R} &= \left(\mathbf{v} \mathbf{v}^H\right) \odot \sum_{n=1}^{N_\mathrm{sub}} \lambda_n \boldsymbol{\vartheta}_n \boldsymbol{\vartheta}_n^H\\
        &= \sum_{n=1}^{N_\mathrm{sub}} \lambda_n \left(\left(\mathbf{v} \odot \boldsymbol{\vartheta}_n\right) \left(\mathbf{v} \odot \boldsymbol{\vartheta}_n\right)^H \right)\\
        &= \sum_{n=1}^{N_\mathrm{sub}} \lambda_n \left(\boldsymbol{\varrho}_n \boldsymbol{\varrho}_n^H\right) \quad \textrm{with} \quad \boldsymbol{\varrho}_n = \mathbf{v} \odot \boldsymbol{\vartheta}_n.
    \end{aligned}
\end{equation}
For usually observed small delay spreads, the time-domain channel realization $\mathbf{\tilde h}_m$ is sparse.
Since $\mathbf{h}_m = \mathbf{F} \mathbf{\tilde h}_m$, where $\mathbf{F} \in \mathbb{C}^{N_\mathrm{sub} \times N_\mathrm{sub}}$ is the \ac{DFT} matrix, $\mathbf{R}_{hh}$ has typically a high spectral gap and therefore, a dominant eigenvector $\boldsymbol{\varrho}_\mathrm{princ}$ exists. Hence, the common spectral pattern $\mathbf{\hat w} = \boldsymbol{\varrho}_\mathrm{princ}$ contains both the signal $\mathbf{v}$ and the common spectral components of the individual physical channel realizations $\mathbf{h}_m$, which might be common signal propagation characteristics, baseband filters or other hardware effects that are mostly constant across datapoints and therefore do not contribute significantly to the \ac{DNN}'s localization.
To keep the portion of common physical signal propagation characteristics, which may contain useful information, as small as possible, a sufficiently large number of spatially distributed receiver antennas is crucial.
If that is ensured, the channel features
\[
    \mathbf{\hat h}_m = \mathbf{o}_m \oslash \mathbf{\hat w} = \left(\mathbf{v} \odot \mathbf{h}_m\right) \oslash \left(\mathbf{v} \odot \boldsymbol{\vartheta}_\mathrm{princ}\right) = \mathbf{h}_m \oslash \boldsymbol{\vartheta}_\mathrm{princ}
\]
still contain unique information about individual signal propagation paths, while the signal $\mathbf{v}$ and mostly redundant information within $\mathbf{h}_m$ is omitted.
Consequently, in a static environment, the obtained channel features $\mathbf{\hat h}_m$ are solely dependent on the \ac{UE} location and therefore meet the aforementioned conditions for reliable \ac{DNN}-based localization.
A visualization of exemplary time-domain realizations of these channel features without obfuscation and with obfuscation, as given in Fig.~\ref{fig:csi_time_domain_RAW_RECOVERED} and Fig.~\ref{fig:csi_time_domain_RECOVERED}, respectively, shows that signal obfuscation has no significant impact on the channel features.

\section{Experimental Results}\label{sec:results}
\begin{table}[h]
    \centering
    \captionof{table}{Scenarios for Training and Evaluation.}
    \vspace{-0.15cm}
    \begin{tabular}{c|c|c|c}
         & \hspace{-0.16cm}\scriptsize{Training ($\mathcal D_\mathrm{train}$)}\hspace{-0.16cm} & \multicolumn{2}{c}{\scriptsize{Evaluation ($\mathcal D_\mathrm{test}$)}}\\ \hline
         \hspace{-0.18cm}\scriptsize{\textbf{Scenario 1}}\hspace{-0.16cm} & \scriptsize{Original CSI} & \scriptsize{\makecell{Original CSI\\(Fig.~\ref{fig:results_fingerprinting_RAW} / \ref{fig:results_cc_RAW})}} & \scriptsize{\makecell{Obfuscated CSI\\(Fig.~\ref{fig:results_fingerprinting_OBFUSCATED} / \ref{fig:results_cc_OBFUSCATED})}}\hspace{-0.16cm}\\ \hline
         \hspace{-0.18cm}\scriptsize{\textbf{Scenario 2}}\hspace{-0.16cm} & \scriptsize{\makecell{Original CSI\\ (recovered)}} & \hspace{-0.16cm}\scriptsize{\makecell{Original CSI (recovered)\\(Fig.~\ref{fig:results_fingerprinting_RAW_RECOVERED} / \ref{fig:results_cc_RAW_RECOVERED})}}\hspace{-0.16cm} & \hspace{-0.16cm}\scriptsize{\makecell{Obfuscated CSI (recovered)\\(Fig.~\ref{fig:results_fingerprinting_RECOVERED} / \ref{fig:results_cc_RECOVERED})}}\hspace{-0.16cm}\\
    \end{tabular}
    \vspace{-0.55cm}
    \label{tab:training_evaluation_scenarios}
\end{table}

\begin{figure*}
    \centering
    \begin{subfigure}[b]{0.24\textwidth}
        \centering
        \begin{tikzpicture}
            \begin{axis}[
                width=0.6\columnwidth,
                height=0.6\columnwidth,
                scale only axis,
                xmin=-12.5,
                xmax=2.5,
                ymin=-14.5,
                ymax=-1.5,
                xlabel = {Coordinate $x_1 ~ [\mathrm{m}]$},
                ylabel = {Coordinate $x_2 ~ [\mathrm{m}]$},
                ylabel shift = -8 pt,
                xlabel shift = -4 pt,
                xtick={-10, -6, -2, 2}
            ]
                \addplot[thick,blue] graphics[xmin=-12.5,ymin=-14.5,xmax=2.5,ymax=-1.5] {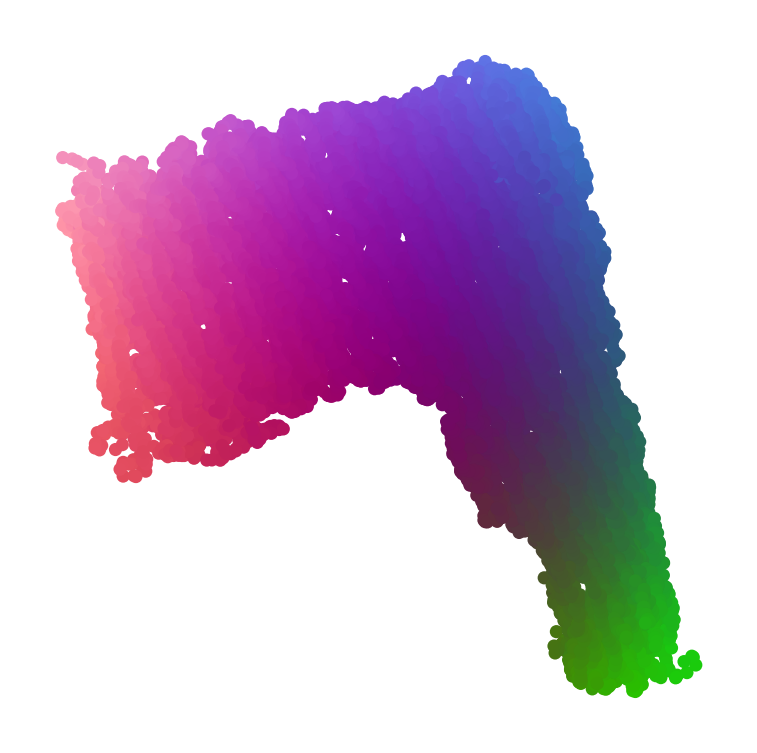};
            \end{axis}
        \end{tikzpicture}
        \vspace{-0.2cm}
        \caption{Original CSI}
        \label{fig:results_fingerprinting_RAW}
    \end{subfigure}
    \begin{subfigure}[b]{0.24\textwidth}
        \centering
        \begin{tikzpicture}
            \begin{axis}[
                width=0.6\columnwidth,
                height=0.6\columnwidth,
                scale only axis,
                xmin=-12.5,
                xmax=2.5,
                ymin=-14.5,
                ymax=-1.5,
                xlabel = {Coordinate $x_1 ~ [\mathrm{m}]$},
                ylabel = {Coordinate $x_2 ~ [\mathrm{m}]$},
                ylabel shift = -8 pt,
                xlabel shift = -4 pt,
                xtick={-10, -6, -2, 2}
            ]
                \addplot[thick,blue] graphics[xmin=-12.5,ymin=-14.5,xmax=2.5,ymax=-1.5] {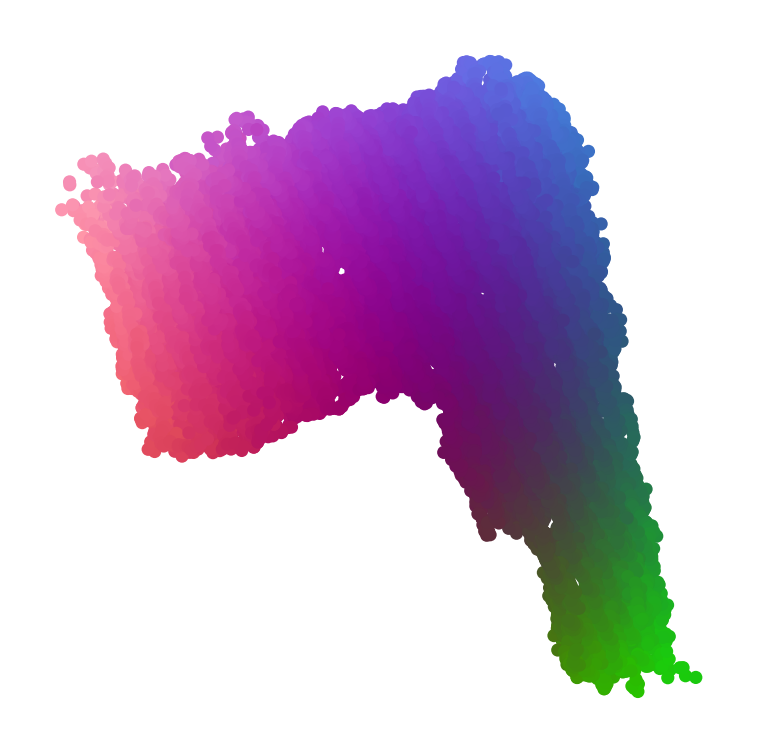};
            \end{axis}
        \end{tikzpicture}
        \vspace{-0.2cm}
        \caption{Original CSI (recovered)}
        \label{fig:results_fingerprinting_RAW_RECOVERED}
    \end{subfigure}
    \begin{subfigure}[b]{0.24\textwidth}
        \centering
        \begin{tikzpicture}
            \begin{axis}[
                width=0.6\columnwidth,
                height=0.6\columnwidth,
                scale only axis,
                xmin=-12.5,
                xmax=2.5,
                ymin=-14.5,
                ymax=-1.5,
                xlabel = {Coordinate $x_1 ~ [\mathrm{m}]$},
                ylabel = {Coordinate $x_2 ~ [\mathrm{m}]$},
                ylabel shift = -8 pt,
                xlabel shift = -4 pt,
                xtick={-10, -6, -2, 2}
            ]
                \addplot[thick,blue] graphics[xmin=-12.5,ymin=-14.5,xmax=2.5,ymax=-1.5] {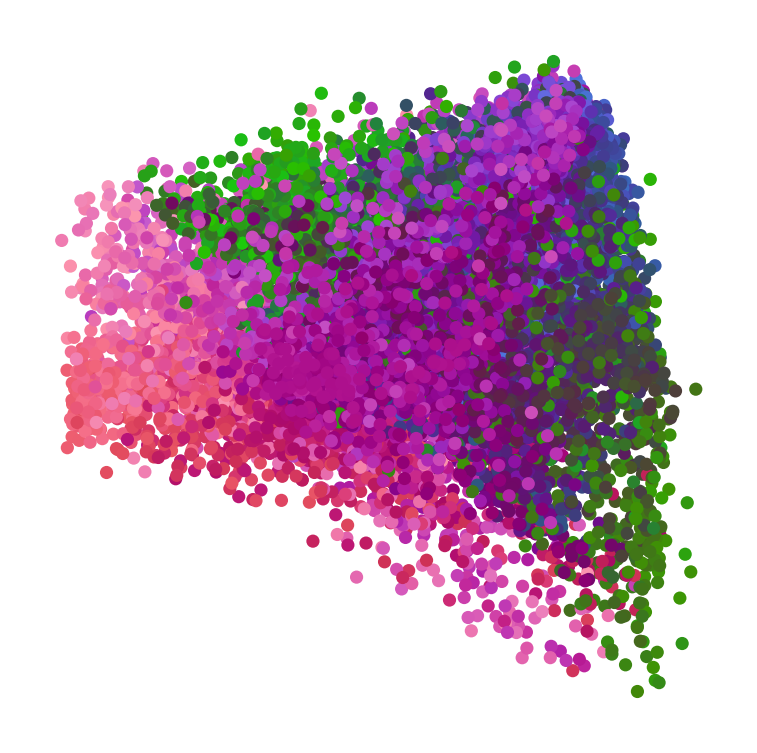};
            \end{axis}
        \end{tikzpicture}
        \vspace{-0.2cm}
        \caption{Obfuscated CSI}
        \label{fig:results_fingerprinting_OBFUSCATED}
    \end{subfigure}
    \begin{subfigure}[b]{0.24\textwidth}
        \centering
        \begin{tikzpicture}
            \begin{axis}[
                width=0.6\columnwidth,
                height=0.6\columnwidth,
                scale only axis,
                xmin=-12.5,
                xmax=2.5,
                ymin=-14.5,
                ymax=-1.5,
                xlabel = {Coordinate $x_1 ~ [\mathrm{m}]$},
                ylabel = {Coordinate $x_2 ~ [\mathrm{m}]$},
                ylabel shift = -8 pt,
                xlabel shift = -4 pt,
                xtick={-10, -6, -2, 2}
            ]
                \addplot[thick,blue] graphics[xmin=-12.5,ymin=-14.5,xmax=2.5,ymax=-1.5] {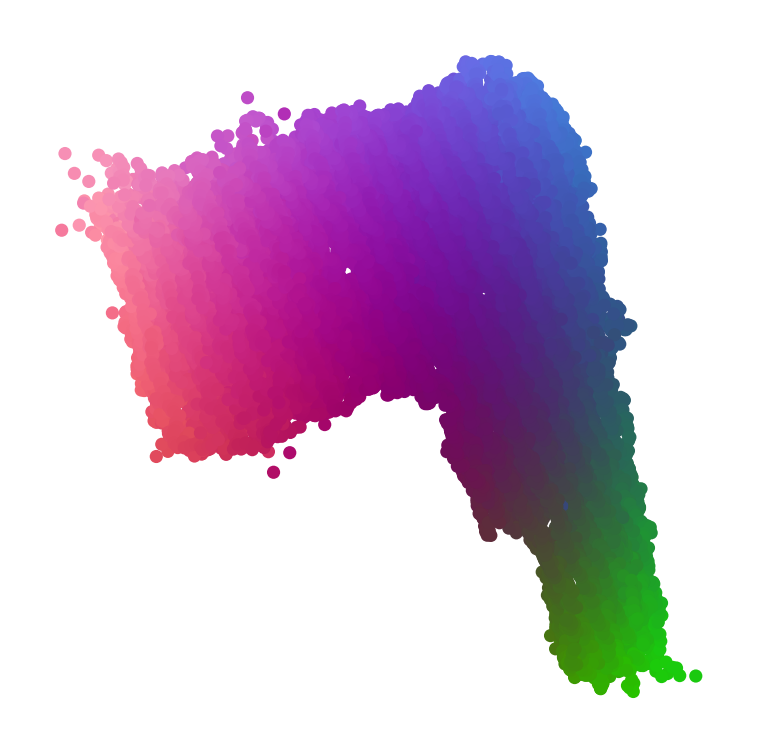};
            \end{axis}
        \end{tikzpicture}
        \vspace{-0.2cm}
        \caption{Obfuscated CSI (recovered)}
        \label{fig:results_fingerprinting_RECOVERED}
    \end{subfigure}
    \vspace{-0.3cm}
    \caption{Position estimates obtained by \ac{CSI} fingerprinting on different channel realizations (color gradient adopted from Fig.~\ref{fig:groundtruth-map}).}
    \vspace{-0.25cm}
    \label{fig:results_fingerprinting}
\end{figure*}

\begin{figure*}
    \centering
    \begin{subfigure}[b]{0.24\textwidth}
        \centering
        \begin{tikzpicture}
            \begin{axis}[
                width=0.6\columnwidth,
                height=0.6\columnwidth,
                scale only axis,
                xmin=-500,
                xmax=300,
                ymin=-300,
                ymax=400,
                xlabel = {Latent variable $z_1$},
                ylabel = {Latent variable $z_2$},
                ylabel shift = -8 pt,
                xlabel shift = -4 pt,
                xtick={-200, 0, 200}
            ]
                \addplot[thick,blue] graphics[xmin=-500,ymin=-300,xmax=300,ymax=400] {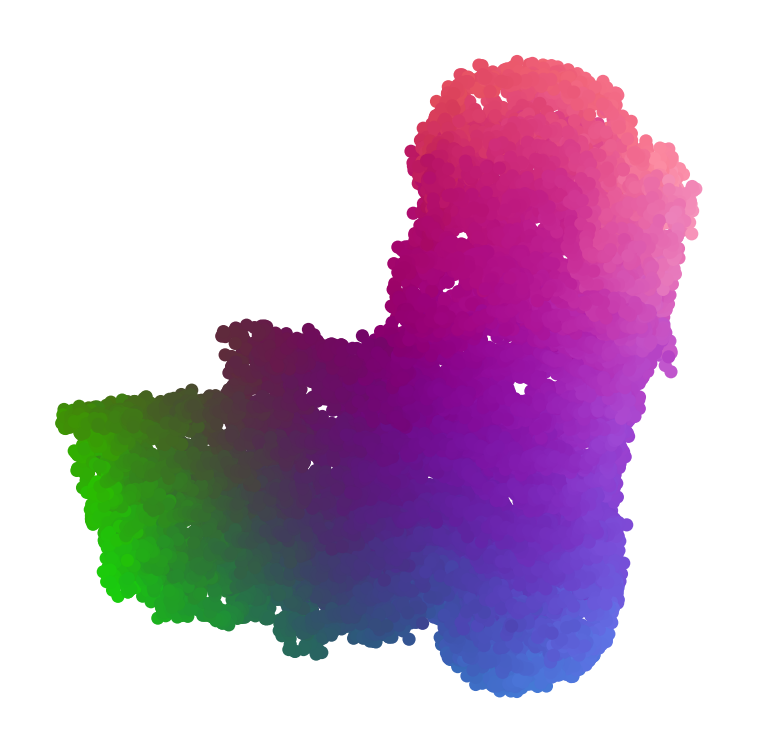};
            \end{axis}
        \end{tikzpicture}
        \vspace{-0.2cm}
        \caption{Original CSI}
        \label{fig:results_cc_RAW}
    \end{subfigure}
    \begin{subfigure}[b]{0.24\textwidth}
        \centering
        \begin{tikzpicture}
            \begin{axis}[
                width=0.6\columnwidth,
                height=0.6\columnwidth,
                scale only axis,
                xmin=-300,
                xmax=450,
                ymin=-350,
                ymax=300,
                xlabel = {Latent variable $z_1$},
                ylabel = {Latent variable $z_2$},
                ylabel shift = -8 pt,
                xlabel shift = -4 pt,
            ]
                \addplot[thick,blue] graphics[xmin=-300,ymin=-350,xmax=450,ymax=300] {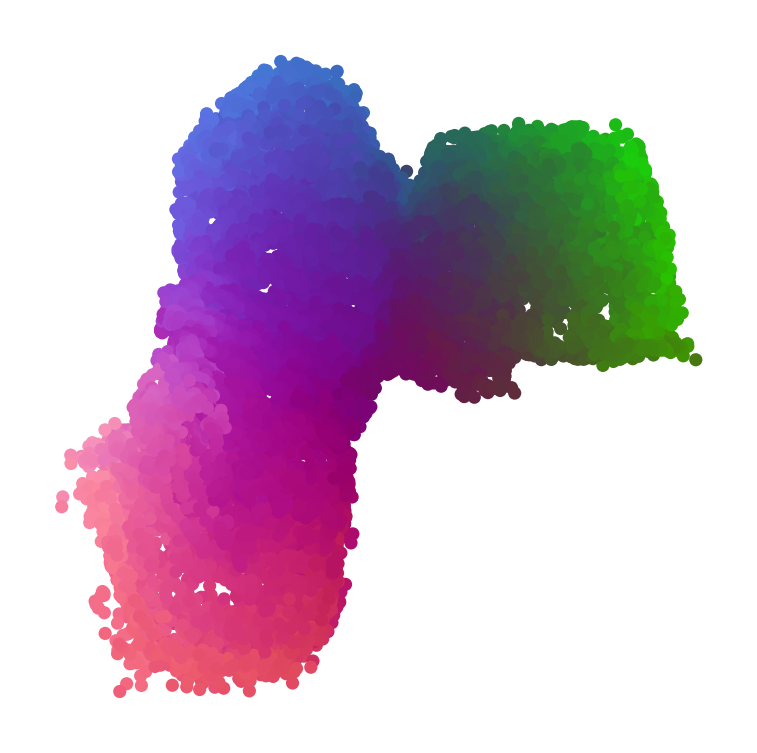};
            \end{axis}
        \end{tikzpicture}
        \vspace{-0.2cm}
        \caption{Original CSI (recovered)}
        \label{fig:results_cc_RAW_RECOVERED}
    \end{subfigure}
    \begin{subfigure}[b]{0.24\textwidth}
        \centering
        \begin{tikzpicture}
            \begin{axis}[
                width=0.6\columnwidth,
                height=0.6\columnwidth,
                scale only axis,
                xmin=-500,
                xmax=300,
                ymin=-350,
                ymax=400,
                xlabel = {Latent variable $z_1$},
                ylabel = {Latent variable $z_2$},
                ylabel shift = -8 pt,
                xlabel shift = -4 pt,
                xtick={-200, 0, 200}
            ]
                \addplot[thick,blue] graphics[xmin=-500,ymin=-350,xmax=300,ymax=400] {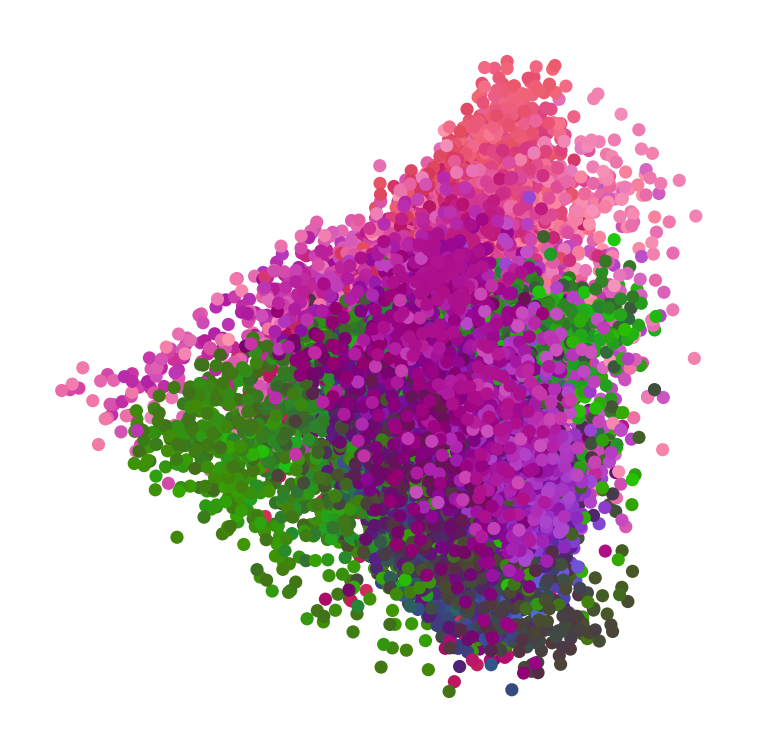};
            \end{axis}
        \end{tikzpicture}
        \vspace{-0.2cm}
        \caption{Obfuscated CSI}
        \label{fig:results_cc_OBFUSCATED}
    \end{subfigure}
    \begin{subfigure}[b]{0.24\textwidth}
        \centering
        \begin{tikzpicture}
            \begin{axis}[
                width=0.6\columnwidth,
                height=0.6\columnwidth,
                scale only axis,
                xmin=-350,
                xmax=450,
                ymin=-350,
                ymax=300,
                xlabel = {Latent variable $z_1$},
                ylabel = {Latent variable $z_2$},
                ylabel shift = -8 pt,
                xlabel shift = -4 pt,
            ]
                \addplot[thick,blue] graphics[xmin=-350,ymin=-350,xmax=450,ymax=300] {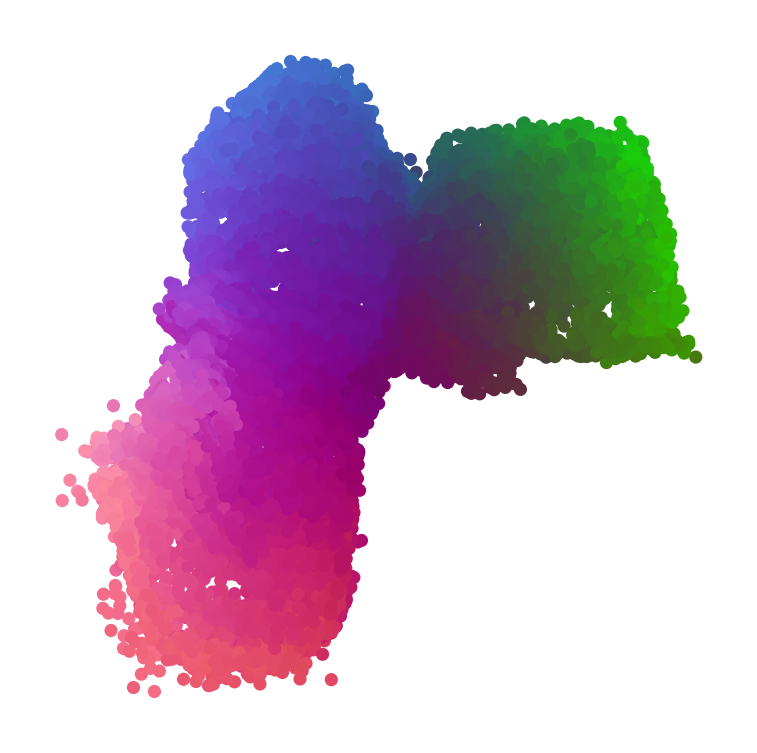};
            \end{axis}
        \end{tikzpicture}
        \vspace{-0.2cm}
        \caption{Obfuscated CSI (recovered)}
        \label{fig:results_cc_RECOVERED}
    \end{subfigure}
    \vspace{-0.3cm}
    \caption{Latent representations obtained by channel charting on different channel realizations (color gradient adopted from Fig.~\ref{fig:groundtruth-map}).}
    \vspace{-0.4cm}
    \label{fig:results_cc}
\end{figure*}

\begin{table*}
    \centering
    \captionof{table}{Evaluation of the \ac{MAE} for different Localization Methods and Antenna Configurations.}
    \vspace{-0.1cm}
    \setlength{\tabcolsep}{4pt}
    \renewcommand{\arraystretch}{0.9}
    \begin{tabular}{l | c | c | c | c | c | c | c}
        \makecell{\textbf{Localization Method}} & \multicolumn{5}{c|}{\textbf{CSI Fingerprinting}} & \textbf{Channel Charting} & \textbf{Triangulation} \\ \hline
        \makecell{\textbf{Antenna Configuration} \\ ($B \times M_\mathrm{r} \times M_\mathrm{c}$) } & $4 \times 2 \times 4$ & $4 \times 2 \times 2$ & $4 \times 1 \times 1$ & \makecell{$2 \times 2 \times 4$ \\ ($b = \{1,3\}$)} & \makecell{$1 \times 2 \times 4$ \\ ($b = 1$)} & $4 \times 2 \times 4$ & $4 \times 2 \times 4$ \\ \hline
        Scenario 1: Original \ac{CSI} & $0.12\,\mathrm m$ & $0.16\,\mathrm m$ & $0.34\,\mathrm m$ & $0.21\,\mathrm m$ & $0.27\,\mathrm m$ & $0.47\,\mathrm m$ & $1.18\,\mathrm m$ \\
        \rowcolor{hellgrau}
        Scenario 1: Obfuscated \ac{CSI} & $3.21\,\mathrm m$ & $3.75\,\mathrm m$ & $7.01\,\mathrm m$ & $6.39\,\mathrm m$ & $4.56\,\mathrm m$ & $2.96\,\mathrm m$ & $1.48\,\mathrm m$ \\ \hline
        Scenario 2: Original \ac{CSI} (recovered) & $0.10\,\mathrm m$ & $0.15\,\mathrm m$ & $0.56\,\mathrm m$ & $0.28\,\mathrm m$ & $0.57\,\mathrm m$ & $0.73\,\mathrm m$ & $1.29\,\mathrm m$ \\
        \rowcolor{hellgrau}
        Scenario 2: Obfuscated \ac{CSI} (recovered) & $0.11\,\mathrm m$ & $0.16\,\mathrm m$ & $0.57\,\mathrm m$ & $0.28\,\mathrm m$ & $0.61\,\mathrm m$ & $0.74\,\mathrm m$ & $1.29\,\mathrm m$ \\
    \end{tabular}
    \label{tab:performance}
    \vspace{-0.4cm}
\end{table*}

The impact of \ac{CSI} obfuscation and the proposed recovery method on the localization performance is evaluated via two scenarios specified in Table~\ref{tab:training_evaluation_scenarios}.
Position estimates obtained by \ac{CSI} fingerprinting are shown in Fig.~\ref{fig:results_fingerprinting}.
For scenario 1, without applying the recovery method (compare Fig.~\ref{fig:results_fingerprinting_RAW} and Fig.~\ref{fig:results_fingerprinting_OBFUSCATED}), \ac{CSI} obfuscation leads to a significant decrease in localization accuracy, which is not observable for scenario 2, where the recovery method is applied (compare Fig.~\ref{fig:results_fingerprinting_RAW_RECOVERED} and Fig.~\ref{fig:results_fingerprinting_RECOVERED}).
Table~\ref{tab:performance}, detailing the \ac{MAE} for the different localization methods and antenna configurations, supports this observation quantitatively.
The middle two rows correspond to scenario 1, the lower two rows to scenario 2.
The recovery step leads to a slight improvement in localization accuracy on the original \ac{CSI} if all \ac{BS} antennas are deployed.
For setups with fewer antenna elements or arrays, there are two observations to be made, namely the heavier disruption of localization accuracy through \ac{CSI} obfuscation in scenario 1, and the increasing negative influence of the recovery step on localization accuracy without \ac{CSI} obfuscation.
However, the latter effect is small in comparison to the gain achieved for the obfuscated \ac{CSI}.
Regardless of the antenna configuration, the localization performance remains mostly unaffected by \ac{CSI} obfuscation with the recovery method (scenario 2).
These results show that the proposed recovery step effectively protects fingerprinting-based localization systems from obfuscating transmitters if a sufficiently large number of \ac{BS} antennas is deployed.

Similar results are observed for channel charting (Fig.~\ref{fig:results_cc}).
Note that channel chart positions typically lie in a transformed version of the physical coordinate frame.
Hence, they are evaluated by computing the \ac{MAE} for the channel chart positions after applying an optimal affine transformation $\mathcal T_\mathrm{opt}(\mathbf{z}^{(l)})$ with respect to the ground truth positions $\mathbf{x}^{(l)}$.
The numerical results are detailed in the second column from the right of Table~\ref{tab:performance}.
The channel charting performance is significantly degraded by \ac{CSI} obfuscation, which is mitigated by the recovery method.
However, it generally suffers notably from the recovery step, which can be explained by hyperparameters of the channel charting algorithm explicitly tuned to the original \ac{CSI}.
We expect that adapted hyperparameters could lead to similar performance as for the original \ac{CSI}.

Nevertheless, channel charting yields significantly better position estimates than classical triangulation, whose results are shown in the rightmost column of Table~\ref{tab:performance}.
The classical localization performance is slightly affected by \ac{CSI} obfuscation and recovery, which can be explained by the delay spread-based heuristic used for the \ac{AoA} likelihood function.

\section{Conclusion and Outlook}\label{sec:conclusion}
We have proposed a \ac{CSI} recovery method that allows multi-antenna \acp{BS} to extract channel features from any (obfuscated) signal transmitted by single-antenna \acp{UE}, enabling adversarial entities to locate these \acp{UE} without their consent.
Therefore, it is crucial to examine further ways to protect the location privacy of users.
Future research can investigate the applicability of the proposed recovery step for multi-antenna \acp{UE} making use of their directivity, and passive object localization.

\bibliographystyle{IEEEtran}
\bibliography{IEEEabrv,references}

\end{document}

%% file: acronyms.tex
\begin{acronym}
 \acro{R2M}{raw 2\textsuperscript{nd} moment}
 \acro{CSI}{channel state information}
 \acro{UE}{user equipment}
 \acro{EM}{electromagnetic}
 \acro{UL}{uplink}
 \acro{BS}{base station}
 \acro{TDD}{time division duplex}
 \acro{FDD}{frequency division duplex}
 \acro{ECC}{error-correcting code}
 \acro{MLD}{maximum likelihood decoding}
 \acro{HDD}{hard decision decoding}
 \acro{IF}{intermediate frequency}
 \acro{RF}{radio frequency}
 \acro{SDD}{soft decision decoding}
 \acro{NND}{neural network decoding}
 \acro{CNN}{convolutional neural network}
 \acro{ML}{maximum likelihood}
 \acro{MLE}{maximum likelihood estimation}
 \acro{GPU}{graphical processing unit}
 \acro{BP}{belief propagation}
 \acro{LTE}{Long Term Evolution}
 \acro{BER}{bit error rate}
 \acro{SNR}{signal-to-noise-ratio}
 \acro{ReLU}{rectified linear unit}
 \acro{BPSK}{binary phase shift keying}
 \acro{QPSK}{quadrature phase shift keying}
 \acro{AWGN}{additive white Gaussian noise}
 \acro{MSE}{mean squared error}
 \acro{LLR}{log-likelihood ratio}
 \acro{MAP}{maximum a posteriori}
 \acro{NVE}{normalized validation error}
 \acro{BCE}{binary cross-entropy}
 \acro{CE}{cross-entropy}
 \acro{BLER}{block error rate}
 \acro{SQR}{signal-to-quantisation-noise-ratio}
 \acro{MIMO}{multiple-input multiple-output}
 \acro{mMIMO}{massive multiple-input multiple-output}
 \acro{OFDM}{orthogonal frequency division multiplex}
 \acro{RF}{radio frequency}
 \acro{LOS}{line of sight}
 \acro{NLoS}{non-line of sight}
 \acro{NMSE}{normalized mean squared error}
 \acro{CFO}{carrier frequency offset}
 \acro{SFO}{sampling frequency offset}
 \acro{IPS}{indoor positioning system}
 \acro{TRIPS}{time-reversal IPS}
 \acro{RSSI}{received signal strength indicator}
 \acro{MIMO}{multiple-input multiple-output}
 \acro{ENoB}{effective number of bits}
 \acro{AGC}{automatic gain control}
 \acro{ADC}{analog to digital converter}
 \acro{ADCs}{analog to digital converters}
 \acro{FB}{front bandpass}
 \acro{FPGA}{field programmable gate array}
 \acro{JSDM}{Joint Spatial Division and Multiplexing}
 \acro{NN}{neural network}
 \acro{IF}{intermediate frequency}
 \acro{LoS}{line-of-sight}
 \acro{NLoS}{non-line-of-sight}
 \acro{DSP}{digital signal processing}
 \acro{AFE}{analog front end}
 \acro{SQNR}{signal-to-quantisation-noise-ratio}
 \acro{SINR}{signal-to-interference-noise-ratio}
 \acro{ENoB}{effective number of bits}
 \acro{PCB}{printed circuit board}
 \acro{EVM}{error vector mangnitude}
 \acro{CDF}{cumulative distribution function}
 \acro{MRC}{maximum ratio combining}
 \acro{MRP}{maximum ratio precoding}
 \acro{MRT}{maximum ratio transmission}
 \acro{DeepL}{deep-learning}
 \acro{DL}{downlink}
 \acro{SISO}{single-input single-output}
 \acro{SGD}{stochastic gradient descent}
 \acro{CP}{cyclic prefix}
 \acro{MISO}{Multiple Input Single Output}
 \acro{LMMSE}{linear minimum mean square error}
 \acro{ZF}{zero forcing}
 \acro{USRP}{universal software radio peripheral}
 \acro{RNN}{recurrent neural network}
 \acro{GRU}{gated recurrent unit}
 \acro{LSTM}{long short-term memory}
 \acro{NTM}{neural turing machine}
 \acro{DNC}{differentiable neural computer}
 \acro{TCN}{temporal convolutional network}
 \acro{FCL}{fully connected layer}
 \acro{MANN}{memory augmented neural network}
 \acro{RNN}{recurrent neural network}
 \acro{DNN}{deep neural network}
 \acro{FIR}{finite impulse response}
 \acro{BPTT}{back-propagation through time}
 \acro{GAN}{generative adversarial network}
 \acro{ELU}{exponential linear unit}
 \acro{tanh}{hyperbolic tangent}
 \acro{BICM}{bit-interleaved coded modulation}
 \acro{OTA}{over-the-air}
 \acro{IM}{intensity modulation}
 \acro{DD}{direct detection}
 \acro{RL}{reinforcement learning}
 \acro{SDR}{software-defined radio}
 \acro{WGAN}{Wasserstein generative adversarial network}
 \acro{BMD}{bit-metric decoding}
 \acro{BMI}{bit-wise mutual information}
 \acro{LDPC}{low-density parity-check}
 \acro{IDD}{iterative demapping and decoding}
 \acro{JSD}{Jensen-Shannon divergence}
 \acro{MMSE}{minimum mean square error}
 \acro{FFT}{fast Fourier transform}
 \acro{DFT}{discrete Fourier transform}
 \acro{IFFT}{inverse fast Fourier transform}
 \acro{QAM}{quadrature amplitude modulation}
 \acro{EMD}{earth mover's distance}
 \acro{TDL}{tapped delay line}
 \acro{KL}{Kullback-Leibler}
 \acro{PRACH}{physical random access channel}
 \acro{URLLC}{ultra-reliable low-latency communication}
 \acro{ANOMA}{asynchronous non-orthogonal multiple access}
 \acro{FEC}{forward error correction}
 \acro{PAPR}{peak-to-average power ratio}
 \acro{APP}{a posteriori probability}
 \acro{COTS}{commercial off-the-shelf}
 \acro{PLL}{phase locked loop}
 \acro{STO}{sampling time offset}
 \acro{SFO}{sampling frequency offset}
 \acro{CFO}{carrier frequency offset}
 \acro{CPO}{carrier phase offset}
 \acro{CSI}{channel state information}
 \acro{GNSS}{global navigation satellite system}
 \acro{ELAA}{extremely large aperture array}
 \acro{UE}{user equipment}
 \acro{DICHASUS}{\underline{Di}stributed \underline{Cha}nnel \underline{S}ounder by \underline{U}niversity of \underline{S}tuttgart}
 \acro{JCaS}{Joint Communication and Sensing}
 \acro{CT}{Continuity}
 \acro{TW}{Trustworthiness}
 \acro{KS}{Kruskal's stress}
 \acro{PCA}{principal component analysis}
 \acro{AoA}{angle of arrival}
 \acro{ToA}{time of arrival}
 \acro{RD}{Rajski's distance}
 \acro{ADP}{angle-delay profile}
 \acro{MPC}{multipath component}
 \acro{CIR}{channel impulse response}
 \acro{MAE}{mean absolute error}
 \acro{MDS}{multidimensional scaling}
 \acro{t-SNE}{t-distributed stochastic neighbor embedding}
 \acro{SM}{Sammon's mapping}
 \acro{CIRA}{channel impulse response amplitude}
 \acro{CS}{cosine similarity}
 \acro{FCF}{forward charting function}
 \acro{CMD}{correlation matrix distance}
 \acro{6G}{Sixth generation}
 \acro{JEPA}{joint-embedding predictive architecture}
 \acro{MAC}{medium access control}
 \acro{WSS}{wide-sense stationary}
\end{acronym}

%% file: corporate_colors.tex
\definecolor{mittelblau}{RGB}{0, 126, 198}
\definecolor{violettblau}{cmyk}{0.9, 0.6, 0, 0}
\definecolor{rot}{RGB}{238, 28 35}
\definecolor{apfelgruen}{RGB}{140, 198, 62}
\definecolor{gelb}{RGB}{1, 221, 0}
\definecolor{orange}{RGB}{244, 111, 33}
\definecolor{pink}{RGB}{237, 0, 140}
\definecolor{lila}{RGB}{128, 10, 145}
\definecolor{hellgrau}{RGB}{224, 224, 224}
\definecolor{mittelgrau}{RGB}{128, 128, 128}
\definecolor{dunkelgrau}{RGB}{80,80,80}
\definecolor{anthrazit}{RGB}{19, 31, 31}

%% file: fig/adversarial_localization_concept.tex
\usetikzlibrary{fit}
\usetikzlibrary{backgrounds}

\begin{tikzpicture}

    \tikzstyle{box1} = [rectangle, draw = dunkelgrau, very thick, rounded corners=1pt,inner sep = 3pt, align = center, fill = dunkelgrau!0!white, minimum height=16pt,minimum width=0.8cm]

    \tikzstyle{box2} = [rectangle, draw = dunkelgrau, very thick, rounded corners=1pt,inner sep = 3pt, align = center, fill = dunkelgrau!0!white, minimum height=1.6cm,minimum width=0.8cm]
    
    \tikzstyle{box3} = [rectangle, draw = dunkelgrau, very thick, rounded corners=1pt,inner sep = 3pt, align = center, minimum height=0.7cm,minimum width=1.6cm]
    
    \tikzstyle{largearoundbox_blue} = [rectangle, draw=mittelblau, very thick, rounded corners=3pt, dashed, inner sep=3pt]
    
    \tikzstyle{largearoundbox_red} = [rectangle, draw=rot, very thick, rounded corners=3pt, dashed, inner sep=3pt]

    
    \node[box1] (TX) at (1,-1.0) {TX};
    \node[box1] (v) at (2.5,-1.0) {$* \mathbf{\tilde v}$};
    
    \node[largearoundbox_blue, fit=(TX) (v)] (Alice) {};
    \node[above=0.1cm of Alice] {Alice (UE)};
    
    \draw [-latex, thick]  (TX.east) -- (v.west)
    node[midway,anchor=south,yshift=-0.05cm]{$\mathbf{\tilde s}$};
    
    \coordinate (antenna_mount_Alice) at (v.east);
    \draw[thick] (antenna_mount_Alice) -- ++(0.3,0) coordinate (antenna_tip_Alice);
    \draw[thick] ([yshift=0.12cm]antenna_tip_Alice) -- ([yshift=-0.12cm]antenna_tip_Alice);

        \draw[semithick] ([xshift=0.1cm]antenna_tip_Alice) ++(0.00,0.1) arc (0:90:0.1cm);
        \draw[semithick] ([xshift=0.15cm]antenna_tip_Alice) ++(0.00,0.1) arc (0:90:0.15cm);
        \draw[semithick] ([xshift=0.2cm]antenna_tip_Alice) ++(0.00,0.1) arc (0:90:0.2cm);

    
    \node[box2] (RX_Bob) at (7.5,-1.0) {RX};

    \node[largearoundbox_blue, fit=(RX_Bob)] (Bob) {};
    \node[right=0.1cm of Bob, align=center] {Bob\\(BS)};
    
    \foreach \yoffset in {-0.6, 0.2, 0.6} { 
        \coordinate (antenna_mount_Bob) at ([yshift=\yoffset cm]RX_Bob.west);
        \draw[thick] (antenna_mount_Bob) -- ++(-0.3,0) coordinate (antenna_tip_Bob); 
        \draw[thick] ([yshift=0.12cm]antenna_tip_Bob) -- ([yshift=-0.12cm]antenna_tip_Bob); 
    }
    \node at ([xshift=-0.15cm, yshift=-0.1cm]Bob.west) {$\vdots$};

    
    \node[box3] (RX_Eve) at (5.0,-3.5) {RX};
    
    \node[largearoundbox_red, fit=(RX_Eve)] (Eve) {};
    \node[right=0.1cm of Eve, align=center] {Eve\\(adversarial BS)};
    
            \foreach \xoffset in {-0.6, -0.2, 0.6} { 
                \coordinate (antenna_mount_Eve) at ([xshift=\xoffset cm]RX_Eve.north);
                \draw[thick] (antenna_mount_Eve) -- ++(0,0.3) coordinate (antenna_tip_Eve); 
                \draw[thick] ([xshift=0.12cm]antenna_tip_Eve) -- ([xshift=-0.12cm]antenna_tip_Eve); 
            }
            \node at ([xshift=0.2cm, yshift=0.15cm]Eve.north) {$\hdots$};


    \draw [-latex, thick]  (antenna_tip_Alice.east) ++ (0.2cm,0.0cm) -- ([xshift=-0.5cm]Bob.west)
    node[midway,anchor=south,yshift=-0.05cm]{$\mathbf{h}_{\mathrm{Bob},m}$};
    
    \draw [-latex, thick]  (antenna_tip_Alice.east) ++ (0.1cm,-0.1cm) -- ([yshift=0.5cm]Eve.north)
    node[midway,anchor=west]{$\mathbf{h}_{\mathrm{Eve},m}$};

    \draw[latex-latex, thick, mittelblau, bend left=20] (Alice.north east) to node[above, midway] {data} (Bob.north west);
    
    \draw[-latex, thick, rot, bend left=20] (Eve.west) to node[below left, midway, align=center] {unauthorized\\localization} (Alice.south);

\end{tikzpicture}

%% file: fig/dnn_structure.tex
\begin{tikzpicture}

\tikzstyle{box1} = [rectangle, draw = mittelblau, thick, rounded corners=1pt,inner sep = 1pt, align = center, fill = mittelblau!10!white,minimum width=4.5cm]

    \node (input) [anchor = south] {$\mathbf f^{(l)}$};
	\node (l1) [box1, below = 0.2cm of input] {Dense, ReLU, 1024 Neurons};
	\node (l2) [box1, below = 0.2cm of l1] {Dense, ReLU, 512 Neurons};
	\node (l3) [box1, below = 0.2cm of l2] {Dense, ReLU, 256 Neurons};
	\node (l4) [box1, below = 0.2cm of l3] {Dense, ReLU, 128 Neurons};
	\node (l5) [box1, below = 0.2cm of l4] {Dense, ReLU, 64 Neurons};
	\node (l6) [box1, below = 0.2cm of l5] {Dense, Linear, 2 Neurons};

	\node (output) [anchor = north] at ($(l6.south) + (0, -0.2)$) {$\hat {\mathbf x}^{(l)}$};

    \draw [-latex] (input) -- (l1);
    \draw [-latex] (l1) -- (l2);
    \draw [-latex] (l2) -- (l3);
    \draw [-latex] (l3) -- (l4);
    \draw [-latex] (l4) -- (l5);
	\draw [-latex] (l5) -- (l6);
	\draw [-latex] (l6) -- (output);
\end{tikzpicture}

%% file: fig/siamese_structure.tex
\begin{tikzpicture}
    
    \tikzstyle{box1} = [rectangle, draw = mittelblau, very thick, rounded corners=1pt,inner sep = 3pt, align = center, fill = mittelblau!10!white, minimum height=16pt,minimum width=1.5cm]
    \tikzstyle{box2} = [rectangle, draw = dunkelgrau, very thick, rounded corners=1pt,inner sep = 3pt, align = center, minimum height=16pt,minimum width=3.5cm]

    \node (in_1) at (2,0){$\mathbf{f}^{(i)}$};
    \node (in_2) at (4,0){$\mathbf{f}^{(j)}$};
    
    \node (in_3) at (3,0){$d_{ij}$};
    
    \node[box1] (dnn_1) at (2,-1.0) {DNN $\mathcal{C}_{\boldsymbol{\theta}}$};
    \node[box1] (dnn_2) at (4,-1.0) {DNN $\mathcal{C}_{\boldsymbol{\theta}}$};
    
    
    \node[box2] (contrastive_loss) at (3,-2.3) {Loss $\mathcal{L}_\mathrm{Siamese}$};
    
    \draw [-latex]  (in_1.south) -- (dnn_1.north)
    node[midway,anchor=east]{};
    \draw [-latex]  (in_2.south) -- (dnn_2.north)
    node[midway,anchor=east]{};
    
    \draw [-latex]  (in_3.south) -- (contrastive_loss)
    node[midway,anchor=east]{};
    
    
    \draw [-latex]  (dnn_1.south) -- (dnn_1.south|-contrastive_loss.north)
    node[midway,anchor=east]{$\mathbf{z}^{(i)}$};
    \draw [-latex]  (dnn_2.south) -- (dnn_2.south|-contrastive_loss.north)
    node[midway,anchor=west]{$\mathbf{z}^{(j)}$};
\end{tikzpicture}

%% file: fig/csi_time_domain_RAW.tex
\begin{tikzpicture}

\definecolor{crimson2143940}{RGB}{214,39,40}
\definecolor{darkgray176}{RGB}{176,176,176}
\definecolor{darkorange25512714}{RGB}{255,127,14}
\definecolor{forestgreen4416044}{RGB}{44,160,44}
\definecolor{gray127}{RGB}{127,127,127}
\definecolor{mediumpurple148103189}{RGB}{148,103,189}
\definecolor{orchid227119194}{RGB}{227,119,194}
\definecolor{sienna1408675}{RGB}{140,86,75}
\definecolor{steelblue31119180}{RGB}{31,119,180}

\begin{axis}[
width=1.25\columnwidth,
height=.6\columnwidth,
tick align=outside,
tick pos=left,
x grid style={darkgray176},
xlabel style={yshift=0.25cm},
xlabel={$\tau$},
xmajorgrids,
xmin=25, xmax=50,
xtick style={color=black},
y grid style={darkgray176},
ylabel={$|\tilde {\mathbf h}_{m}^{(l)}|$},
ymajorgrids,
ymin=-0.15, ymax=2.65,
ytick style={color=black},
ticklabel style={fill=white},
legend cell align={left},
legend style={
  at={(0.77,0.96)},
  anchor=north,
},
legend columns=2,
clip = false]
\path [draw=none, fill=red, fill opacity=0.15]
(axis cs:27,-0.15)
--(axis cs:27,2.65) node (taumintop) {}
--(axis cs:40,2.65) node (taumaxtop) {}
--(axis cs:40,-0.15)
--cycle;

\addplot [semithick, mittelblau]
table {%
25 0.0223980577003032
26 0.0280409742318143
27 0.0296968359991254
28 0.036407469082542
29 0.0401608993614583
30 0.0742981684272953
31 1.76215277993609
32 1.43569068918462
33 0.626221778044929
34 0.35786200173976
35 0.249071418610153
36 0.324418986101388
37 0.21014335476078
38 0.132431219134516
39 0.052933183138878
40 0.241013661810374
41 0.0562645025006037
42 0.211430724845306
43 0.0408007403266638
44 0.103765962270633
45 0.123815165933343
46 0.110289646892119
47 0.13903972052481
48 0.0975793358108486
49 0.0379371425698216
50 0.172890195853465
};
\addlegendentry{$m = 1$}
\addplot [semithick, darkorange25512714]
table {%
25 0.0235526682916398
26 0.0293602430894197
27 0.0329711199014191
28 0.0439674590636246
29 0.056415623638642
30 0.121557826201665
31 1.22411919671112
32 1.26673424414638
33 0.378116975488814
34 0.369437392134214
35 0.219749561347063
36 0.0752198762621541
37 0.167857104533353
38 0.11780345052302
39 0.133275362862565
40 0.0897945375472251
41 0.0688828419422661
42 0.0935696433163136
43 0.225608771070459
44 0.0959040042909776
45 0.0824678415961891
46 0.0764619792802313
47 0.0254191214326955
48 0.0831850150283149
49 0.088195147207263
50 0.0793738677763425
};
\addlegendentry{$m = 2$}
\addplot [semithick, forestgreen4416044]
table {%
25 0.0227947833005773
26 0.0276894431191414
27 0.0292348257653876
28 0.035373407653441
29 0.0329673250996796
30 0.0433759434756096
31 1.58479359764561
32 1.23346595023839
33 0.249749538369589
34 0.329154743379769
35 0.322760035955224
36 0.070917944666153
37 0.262895548620649
38 0.325771743367839
39 0.13430984389164
40 0.208278155594557
41 0.121466981127254
42 0.146160110095744
43 0.126766602715954
44 0.153465188344799
45 0.0577081652094731
46 0.144077217294065
47 0.077060670130291
48 0.148735715335197
49 0.138935236487061
50 0.03031280683082
};
\addlegendentry{$m = 3$}
\addplot [semithick, crimson2143940]
table {%
25 0.00800089614811418
26 0.0107901762527008
27 0.0127727003135537
28 0.0219678177124036
29 0.070211366696218
30 0.367029394543282
31 2.08936961206349
32 0.498729164976987
33 0.388510234986395
34 0.351188695212273
35 0.307406868723049
36 0.138741867264417
37 0.0797417166982415
38 0.102047562967868
39 0.0637989057911018
40 0.282715757720877
41 0.163906537268558
42 0.00857431468245911
43 0.0329390989369477
44 0.0531626033021163
45 0.0553051567319261
46 0.182288504932392
47 0.0821028901144985
48 0.0541194153953792
49 0.0186534759901473
50 0.0389032126280866
};
\addlegendentry{$m = 4$}
\addplot [semithick, mediumpurple148103189]
table {%
25 0.0169802814868508
26 0.0224006422910729
27 0.0240068411633501
28 0.0336717117849213
29 0.0462294162281025
30 0.110590143006527
31 1.11735511180322
32 1.3205915508305
33 0.263316295525919
34 0.409576196844635
35 0.188563460016423
36 0.290255915130973
37 0.207079546715818
38 0.176470381154976
39 0.0748981368086964
40 0.123316578956211
41 0.0868759060745405
42 0.243667414369905
43 0.171825711046363
44 0.159080061727429
45 0.237382343941831
46 0.0743566719166925
47 0.0472903797083276
48 0.0866879142275797
49 0.0515668743356483
50 0.149749710394555
};
\addlegendentry{$m = 5$}
\addplot [semithick, sienna1408675]
table {%
25 0.0140569043930323
26 0.0171890062122608
27 0.0153744919828879
28 0.0133440675738466
29 0.00702270680013649
30 0.0777372090473524
31 1.29797715570232
32 0.532049326790323
33 0.405118401937614
34 0.400507642479597
35 0.0288120275713148
36 0.308662874179224
37 0.0738364147197365
38 0.106612541364425
39 0.241842785354532
40 0.11925805883006
41 0.0811047343857697
42 0.186306024569002
43 0.0791315601255294
44 0.169324297411967
45 0.171973717111328
46 0.0906155572918978
47 0.138293579549569
48 0.0887532307782234
49 0.0355926489720129
50 0.12641945348855
};
\addlegendentry{$m = 6$}
\addplot [semithick, orchid227119194]
table {%
25 0.0198642713390635
26 0.0246340654495543
27 0.0264595797800962
28 0.0329516258864267
29 0.034616895505258
30 0.0541259024612909
31 1.43618154377265
32 1.20405298666577
33 0.21573532129307
34 0.447151074264336
35 0.273778542582415
36 0.396827693072208
37 0.319074712585491
38 0.105960439848436
39 0.20880402411518
40 0.147942256605627
41 0.204969168215089
42 0.16005071079064
43 0.134857456140869
44 0.160596982432444
45 0.173958079642936
46 0.184626979570817
47 0.08760113397418
48 0.119192593027602
49 0.0661578958091527
50 0.0610134681420758
};
\addlegendentry{$m = 7$}
\addplot [semithick, gray127]
table {%
25 0.00930320948755469
26 0.0141868966216171
27 0.0172331707997872
28 0.0274253299343743
29 0.0467039964817883
30 0.130290773224405
31 0.862542013857823
32 1.8646043211185
33 0.176701628440154
34 0.309984451908497
35 0.105309493668171
36 0.369882357942025
37 0.163360319396303
38 0.0646546407240265
39 0.120688777985011
40 0.208692248592672
41 0.24734862554196
42 0.0519095171377572
43 0.157088709036792
44 0.130561622210939
45 0.11702631132531
46 0.074018962688883
47 0.12037579121536
48 0.139466800445059
49 0.0699054526010555
50 0.0783663741987024
};
\addlegendentry{$m = 8$}

\draw [thick, dotted] (axis cs: 27, 2.65) -- (axis cs: 27, -0.15) node[pos = 0.65, anchor = center, fill = white, opacity = 0.7, text opacity = 1.0, inner sep = 1pt, red!80!black, fill = white] {$\tau_\mathrm{min}$};
\draw [thick, dotted] (axis cs: 40, 2.65) -- (axis cs: 40, -0.15) node[pos = 0.65, anchor = center, fill = white, opacity = 0.7, text opacity = 1.0, inner sep = 1pt, red!80!black, fill = white] {$\tau_\mathrm{max}$};
\end{axis}

\end{tikzpicture}

%% file: fig/csi_time_domain_RAW_RECOVERED.tex
\begin{tikzpicture}

\definecolor{crimson2143940}{RGB}{214,39,40}
\definecolor{darkgray176}{RGB}{176,176,176}
\definecolor{darkorange25512714}{RGB}{255,127,14}
\definecolor{forestgreen4416044}{RGB}{44,160,44}
\definecolor{gray127}{RGB}{127,127,127}
\definecolor{mediumpurple148103189}{RGB}{148,103,189}
\definecolor{orchid227119194}{RGB}{227,119,194}
\definecolor{sienna1408675}{RGB}{140,86,75}
\definecolor{steelblue31119180}{RGB}{31,119,180}

\begin{axis}[
width=1.25\columnwidth,
height=.6\columnwidth,
tick align=outside,
tick pos=left,
x grid style={darkgray176},
xlabel style={yshift=0.25cm},
xlabel={$\tau$},
xmajorgrids,
xmin=25, xmax=50,
xtick style={color=black},
y grid style={darkgray176},
ylabel={$|\tilde {\hat{\mathbf{h}}}_{\mathrm{original},m}^{(l)}|$},
ymajorgrids,
ymin=-0.15, ymax=2.65,
ytick style={color=black},
ticklabel style={fill=white},
legend cell align={left},
legend style={
  at={(0.77,0.96)},
  anchor=north,
},
legend columns=2,
clip = false]
\path [draw=none, fill=red, fill opacity=0.15]
(axis cs:27,-0.15)
--(axis cs:27,2.65) node (taumintop) {}
--(axis cs:40,2.65) node (taumaxtop) {}
--(axis cs:40,-0.15)
--cycle;

\addplot [semithick, mittelblau]
table {%
25 0.0642220355090363
26 0.0650845727450505
27 0.0670841961405385
28 0.0702095650472546
29 0.075988461466216
30 0.0901295182471773
31 0.145486232732645
32 1.11777308138552
33 0.0871427850293262
34 0.165476839031157
35 0.184572321849345
36 0.172638297784017
37 0.139942923825885
38 0.0234881506284538
39 0.0645866290368297
40 0.121102128876067
41 0.15988878976083
42 0.15723002462
43 0.0253658855957454
44 0.0356119430404401
45 0.0119966907984583
46 0.0718211746668503
47 0.0731306187284192
48 0.0685711229201518
49 0.0792067440681888
50 0.0651395630634264
};
\addlegendentry{$m = 1$}
\addplot [semithick, darkorange25512714]
table {%
25 0.0494866423694724
26 0.0514284979927389
27 0.0538534484941803
28 0.0583106686310063
29 0.0670462051620494
30 0.0876743444497679
31 0.161757170134379
32 0.835925839666768
33 0.0945508620537603
34 0.166149386920571
35 0.0863975935190691
36 0.140373578798856
37 0.112004246063853
38 0.0777946969897633
39 0.0781934543733205
40 0.034182155766807
41 0.0435830089541161
42 0.0625786398372193
43 0.102704380316605
44 0.120096142028261
45 0.0499844168210149
46 0.0325140701376542
47 0.0127306120675263
48 0.0231541340867904
49 0.0561822531997575
50 0.0301197657488432
};
\addlegendentry{$m = 2$}
\addplot [semithick, forestgreen4416044]
table {%
25 0.0395167038972419
26 0.0403984067378917
27 0.0412224243571411
28 0.0427852728925413
29 0.0457284152616736
30 0.0531529277994424
31 0.089438991407191
32 0.986453263392242
33 0.0711005434336716
34 0.0738665336994256
35 0.0396888956950139
36 0.165705983722946
37 0.101596388103133
38 0.0804316352447946
39 0.13587964580601
40 0.0501634178307707
41 0.0485427459719336
42 0.0505972070437582
43 0.0469217746238487
44 0.0398624110783526
45 0.124403645036088
46 0.0799062019125065
47 0.030719852279279
48 0.0872838281914812
49 0.0972176775668724
50 0.0560994977128636
};
\addlegendentry{$m = 3$}
\addplot [semithick, crimson2143940]
table {%
25 0.111953168415383
26 0.115851541897621
27 0.122359532653901
28 0.133226420592318
29 0.154613382544415
30 0.206070832289378
31 0.381881808304135
32 0.894184274994911
33 0.35542400499239
34 0.142179831268262
35 0.0644676815645927
36 0.138010187365313
37 0.104703989770864
38 0.130822608096834
39 0.140270623201894
40 0.150110171628006
41 0.119185404552186
42 0.123723946015352
43 0.102614019287372
44 0.128442410156803
45 0.0870354145166206
46 0.0757589017957667
47 0.142925552179005
48 0.102692551510881
49 0.106227944175783
50 0.102564210225072
};
\addlegendentry{$m = 4$}
\addplot [semithick, mediumpurple148103189]
table {%
25 0.0885276921329829
26 0.0909787881113018
27 0.0951084484679874
28 0.101504706402938
29 0.11277894955216
30 0.137124152263916
31 0.211154959660908
32 0.818947012206988
33 0.0898095868318183
34 0.109264850739302
35 0.187655203311306
36 0.104106885498048
37 0.21139125654051
38 0.0648943995983526
39 0.0607510024228669
40 0.0256076548664541
41 0.086033367381581
42 0.144023050187374
43 0.0582374277653192
44 0.0723491656214736
45 0.0888578582819782
46 0.15502437366056
47 0.0779542020033925
48 0.100259522978238
49 0.128005446166684
50 0.114856835408705
};
\addlegendentry{$m = 5$}
\addplot [semithick, sienna1408675]
table {%
25 0.031141065440609
26 0.0341047819316887
27 0.0373842403342535
28 0.0435376260517578
29 0.0543718865112772
30 0.0744812785302081
31 0.113241803324284
32 0.64251678998726
33 0.27996137046145
34 0.175384518790524
35 0.118067195764768
36 0.079710841250097
37 0.191241007820697
38 0.110237665308737
39 0.108690812238059
40 0.0277997833632554
41 0.0775942798642444
42 0.0321774743255669
43 0.0721120126420611
44 0.0188549636813052
45 0.0800828686145349
46 0.0832065674549068
47 0.0551268403842788
48 0.0628430569156226
49 0.0243106736773241
50 0.0188480640488157
};
\addlegendentry{$m = 6$}
\addplot [semithick, orchid227119194]
table {%
25 0.0115845895129198
26 0.0122350843163394
27 0.0128709989243521
28 0.0144981395373226
29 0.0170391954737288
30 0.0237340958690513
31 0.0608737412405492
32 0.868760796776699
33 0.0556264606149904
34 0.0631994020028334
35 0.128979582004814
36 0.0871167980674489
37 0.218106101242315
38 0.11535293465726
39 0.112093704999159
40 0.0656611264926089
41 0.0453728312704388
42 0.0720901752655642
43 0.147460009691345
44 0.0279179670033364
45 0.077345207331848
46 0.104358929100753
47 0.0326762429006736
48 0.040972000502991
49 0.0617968542691204
50 0.00661104519852141
};
\addlegendentry{$m = 7$}
\addplot [semithick, gray127]
table {%
25 0.102317933559251
26 0.104112505353798
27 0.107456195241572
28 0.113938089585796
29 0.127133078489955
30 0.157068228358116
31 0.244539572054522
32 0.707501187907188
33 0.49744411636225
34 0.241580710733038
35 0.216715406779217
36 0.192624659452066
37 0.110331012677087
38 0.143456159274106
39 0.0772133015547947
40 0.0142024461464468
41 0.155925280128036
42 0.0597775056529045
43 0.0885421462154376
44 0.151157159884525
45 0.138188814508236
46 0.173983966623429
47 0.126635491883454
48 0.0842880519548797
49 0.12433863136519
50 0.143577319636355
};
\addlegendentry{$m = 8$}

\draw [thick, dotted] (axis cs: 27, 2.65) -- (axis cs: 27, -0.15) node[pos = 0.65, anchor = center, fill = white, opacity = 0.7, text opacity = 1.0, inner sep = 1pt, red!80!black, fill = white] {$\tau_\mathrm{min}$};
\draw [thick, dotted] (axis cs: 40, 2.65) -- (axis cs: 40, -0.15) node[pos = 0.65, anchor = center, fill = white, opacity = 0.7, text opacity = 1.0, inner sep = 1pt, red!80!black, fill = white] {$\tau_\mathrm{max}$};
\end{axis}

\end{tikzpicture}

%% file: fig/csi_time_domain_OBFUSCATED.tex
\begin{tikzpicture}

\definecolor{crimson2143940}{RGB}{214,39,40}
\definecolor{darkgray176}{RGB}{176,176,176}
\definecolor{darkorange25512714}{RGB}{255,127,14}
\definecolor{forestgreen4416044}{RGB}{44,160,44}
\definecolor{gray127}{RGB}{127,127,127}
\definecolor{mediumpurple148103189}{RGB}{148,103,189}
\definecolor{orchid227119194}{RGB}{227,119,194}
\definecolor{sienna1408675}{RGB}{140,86,75}
\definecolor{steelblue31119180}{RGB}{31,119,180}

\begin{axis}[
width=1.25\columnwidth,
height=.6\columnwidth,
tick align=outside,
tick pos=left,
x grid style={darkgray176},
xlabel style={yshift=0.25cm},
xlabel={$\tau$},
xmajorgrids,
xmin=25, xmax=50,
xtick style={color=black},
y grid style={darkgray176},
ylabel={$|\tilde {\mathbf o}_{m}^{(l)}|$},
ymajorgrids,
ymin=-0.15, ymax=2.65,
ytick style={color=black},
ticklabel style={fill=white},
legend cell align={left},
legend style={
  at={(0.77,0.96)},
  anchor=north,
},
legend columns=2,
clip = false]
\path [draw=none, fill=red, fill opacity=0.15]
(axis cs:27,-0.15)
--(axis cs:27,2.65) node (taumintop) {}
--(axis cs:40,2.65) node (taumaxtop) {}
--(axis cs:40,-0.15)
--cycle;

\addplot [semithick, mittelblau]
table {%
25 0.00480243753012096
26 0.00758326232008515
27 0.0063279447922761
28 0.00885890755373678
29 0.0084276232738819
30 0.0204477841706729
31 0.596558313890555
32 0.498459276108656
33 0.37311138364624
34 0.394068183167207
35 0.333079874242385
36 0.392087283278955
37 0.732535509720922
38 1.16005930084657
39 0.162688224175042
40 0.564028792928366
41 0.208781788099553
42 0.544027849842175
43 0.62838505760087
44 0.51691148976179
45 0.5912896534032
46 0.863427100589409
47 0.54930888220876
48 0.294313459850769
49 0.434771766328524
50 0.151590582995968
};
\addlegendentry{$m = 1$}
\addplot [semithick, darkorange25512714]
table {%
25 0.00551681813639699
26 0.00802961761759164
27 0.00712315059765958
28 0.0114379634707165
29 0.0136626565487552
30 0.0352328004497613
31 0.409455615754684
32 0.449467306929929
33 0.139636102019732
34 0.268987734244167
35 0.34884074396798
36 0.358507422653859
37 0.591219964584043
38 0.966381355171512
39 0.199414846408246
40 0.368054863134991
41 0.305463654095115
42 0.326894775522113
43 0.402307317386231
44 0.678116778615439
45 0.707557983307527
46 0.855462961974782
47 0.490450378475372
48 0.0950686203784788
49 0.0736024437699997
50 0.0203555063635809
};
\addlegendentry{$m = 2$}
\addplot [semithick, forestgreen4416044]
table {%
25 0.00466474956328166
26 0.00719783520628405
27 0.00636489408463315
28 0.00846803186595461
29 0.00548020506477088
30 0.0103944181743276
31 0.534567987382855
32 0.421290781747796
33 0.245301193106124
34 0.3376756574497
35 0.416115057407119
36 0.448574511438232
37 0.685216219942183
38 1.18258124054371
39 0.355842994672194
40 0.626246179280782
41 0.191164277363788
42 0.289213863489455
43 0.382372759245731
44 0.46041648384755
45 0.654722382072441
46 0.701245735476902
47 0.324273693549719
48 0.226708722677446
49 0.110702950496255
50 0.095074107298717
};
\addlegendentry{$m = 3$}
\addplot [semithick, crimson2143940]
table {%
25 0.000938500678740969
26 0.00370208163680966
27 0.00312340542373795
28 0.00713981722662364
29 0.0237969777128929
30 0.124016790858543
31 0.716741951191184
32 0.137600946081717
33 0.227945505496068
34 0.446437567692976
35 0.587018668972787
36 0.345055008407012
37 0.946825521864966
38 0.929842021427248
39 0.605072272902858
40 0.465154805292049
41 0.191923088649779
42 0.415079628127213
43 0.620958301479324
44 0.752376253479759
45 0.641781304818575
46 0.602253664185014
47 0.102599803631027
48 0.156637642187096
49 0.101062641359273
50 0.121398709611914
};
\addlegendentry{$m = 4$}
\addplot [semithick, mediumpurple148103189]
table {%
25 0.00410578757189382
26 0.00629750669386064
27 0.00562371085028722
28 0.00926888709119953
29 0.0118578174766452
30 0.0332096750470442
31 0.375526017425832
32 0.462367768494002
33 0.202333857690906
34 0.253355090610363
35 0.294600948875007
36 0.349782334536797
37 0.517378688053264
38 1.026841804823
39 0.202858671012111
40 0.372980771050495
41 0.228993705027645
42 0.273953494923788
43 0.477713369818325
44 0.406499586745338
45 0.597110412276147
46 0.659081342779449
47 0.443528189905558
48 0.321999498216889
49 0.13778576972053
50 0.0473383934373237
};
\addlegendentry{$m = 5$}
\addplot [semithick, sienna1408675]
table {%
25 0.00304686093625389
26 0.00475385352391057
27 0.00260820976613026
28 0.003143613049487
29 0.00546884275294825
30 0.0283232418501623
31 0.443795356917605
32 0.183094981532858
33 0.103282477016964
34 0.295349233928465
35 0.259885128871743
36 0.350214387194428
37 0.594736395236882
38 0.591988505323274
39 0.169353702541393
40 0.309900707196938
41 0.172594465106833
42 0.235145816055572
43 0.404177612198014
44 0.676172709242402
45 0.810458078780932
46 0.818219151933147
47 0.30103149158412
48 0.188096573747054
49 0.240785734871452
50 0.0902855969493358
};
\addlegendentry{$m = 6$}
\addplot [semithick, orchid227119194]
table {%
25 0.00424229762155975
26 0.00646147635550891
27 0.00589117087780552
28 0.00785083743338799
29 0.00716569528752999
30 0.0138733748338583
31 0.483427914393265
32 0.410575491714681
33 0.193336037835735
34 0.238898967586618
35 0.387645765966379
36 0.54143808613408
37 0.610021933056838
38 1.00459597928637
39 0.371288670521265
40 0.634316880550725
41 0.214102453081944
42 0.330148772874927
43 0.225050737981856
44 0.768508477324333
45 0.882814123529865
46 0.742433215487798
47 0.50709610382932
48 0.185265258388791
49 0.239349026281457
50 0.168564102743328
51 0.0960623408814318
};
\addlegendentry{$m = 7$}
\addplot [semithick, gray127]
table {%
25 0.00229217160073362
26 0.00340886778529117
27 0.0042100661145703
28 0.00777547847410537
29 0.0129698433504918
30 0.0411246949937543
31 0.289004507193511
32 0.646958537104084
33 0.132165991999501
34 0.123705813907704
35 0.3743419824832
36 0.617215717648912
37 0.380999060179662
38 1.00170227190757
39 0.566251879959054
40 0.59697693377235
41 0.377812813103318
42 0.0679398363189499
43 0.241327187854593
44 0.643292885695558
45 0.822685436732927
46 0.542661282755224
47 0.670422209414183
48 0.154984665839666
49 0.20389686593144
50 0.154770343028893
};
\addlegendentry{$m = 8$}

\draw [thick, dotted] (axis cs: 27, 2.65) -- (axis cs: 27, -0.15) node[pos = 0.65, anchor = center, fill = white, opacity = 0.7, text opacity = 1.0, inner sep = 1pt, red!80!black, fill = white] {$\tau_\mathrm{min}$};
\draw [thick, dotted] (axis cs: 40, 2.65) -- (axis cs: 40, -0.15) node[pos = 0.65, anchor = center, fill = white, opacity = 0.7, text opacity = 1.0, inner sep = 1pt, red!80!black, fill = white] {$\tau_\mathrm{max}$};
\end{axis}

\end{tikzpicture}

%% file: fig/csi_time_domain_RECOVERED.tex
\begin{tikzpicture}

\definecolor{crimson2143940}{RGB}{214,39,40}
\definecolor{darkgray176}{RGB}{176,176,176}
\definecolor{darkorange25512714}{RGB}{255,127,14}
\definecolor{forestgreen4416044}{RGB}{44,160,44}
\definecolor{gray127}{RGB}{127,127,127}
\definecolor{mediumpurple148103189}{RGB}{148,103,189}
\definecolor{orchid227119194}{RGB}{227,119,194}
\definecolor{sienna1408675}{RGB}{140,86,75}
\definecolor{steelblue31119180}{RGB}{31,119,180}

\begin{axis}[
width=1.25\columnwidth,
height=.6\columnwidth,
tick align=outside,
tick pos=left,
x grid style={darkgray176},
xlabel style={yshift=0.25cm},
xlabel={$\tau$},
xmajorgrids,
xmin=25, xmax=50,
xtick style={color=black},
y grid style={darkgray176},
ylabel={$|\tilde {\hat{\mathbf{h}}}_{\mathrm{obfuscated},m}^{(l)}|$},
ymajorgrids,
ymin=-0.15, ymax=2.65,
ytick style={color=black},
ticklabel style={fill=white},
legend cell align={left},
legend style={
  at={(0.77,0.96)},
  anchor=north,
},
legend columns=2,
clip = false]
\path [draw=none, fill=red, fill opacity=0.15]
(axis cs:27,-0.15)
--(axis cs:27,2.65) node (taumintop) {}
--(axis cs:40,2.65) node (taumaxtop) {}
--(axis cs:40,-0.15)
--cycle;

\addplot [semithick, mittelblau]
table {%
25 0.0646904948383502
26 0.0651923448488591
27 0.0668306778806528
28 0.0694858886842166
29 0.0747362132458571
30 0.0884212606295138
31 0.145637119503709
32 1.12507366490738
33 0.0880835554862011
34 0.176418557720326
35 0.193515003988337
36 0.177345544925115
37 0.152756394899375
38 0.0228097892201276
39 0.0691511760287999
40 0.128924672965121
41 0.168065630270532
42 0.162628382223393
43 0.0287811573580345
44 0.0377504023266023
45 0.00782564417528474
46 0.0761417742821721
47 0.0724922303894476
48 0.0723783911734399
49 0.0850940795150093
50 0.0706182126479356
};
\addlegendentry{$m = 1$}
\addplot [semithick, darkorange25512714]
table {%
25 0.0511300658839844
26 0.0527546676706493
27 0.0549092775954352
28 0.0589578956548188
29 0.0673033544889978
30 0.0878841084861978
31 0.164058339620497
32 0.840957737809279
33 0.0868600499883904
34 0.166497118332226
35 0.0924493593889726
36 0.142531664346964
37 0.113675015787371
38 0.084795101039644
39 0.074930074321413
40 0.0386322202828934
41 0.0519098624040304
42 0.0671885497506441
43 0.109501816285538
44 0.119940305656862
45 0.052360371879981
46 0.0381530884722343
47 0.0175628413683737
48 0.0266782503449486
49 0.0615207859088529
50 0.0287498462500905
};
\addlegendentry{$m = 2$}
\addplot [semithick, forestgreen4416044]
table {%
25 0.0405051963102247
26 0.0413790023898004
27 0.0421569274949167
28 0.0437143184251386
29 0.0468134004760686
30 0.055005183743263
31 0.095820766600999
32 0.998177939924616
33 0.0843032046785057
34 0.0685902067742017
35 0.0372672343001914
36 0.167679774654141
37 0.0977456990167652
38 0.0747052756074201
39 0.141501020140789
40 0.0472738120698205
41 0.0542420000281497
42 0.0516693683297056
43 0.0450726001883365
44 0.0440816122290079
45 0.130433003876682
46 0.0821782061645391
47 0.0366873707368792
48 0.0901625548041125
49 0.0956636408916541
50 0.0597751653387873
};
\addlegendentry{$m = 3$}
\addplot [semithick, crimson2143940]
table {%
25 0.11035397244769
26 0.113731213633428
27 0.119527359520484
28 0.1295984593954
29 0.150083212225568
30 0.200724311676018
31 0.372825314338768
32 0.914092069617972
33 0.381080919453843
34 0.15540984614922
35 0.0754232685909495
36 0.14180682397454
37 0.104451481159579
38 0.132451423713703
39 0.145959476856418
40 0.156790047134912
41 0.114388927393479
42 0.13273768841801
43 0.109842143081176
44 0.133693151530627
45 0.0919871910228444
46 0.0775829000752695
47 0.139040006060288
48 0.104888975642064
49 0.111273742634933
50 0.10535381948722
};
\addlegendentry{$m = 4$}
\addplot [semithick, mediumpurple148103189]
table {%
25 0.0893839065185748
26 0.0914755741220255
27 0.0951971646998568
28 0.101083947661556
29 0.1118687668634
30 0.135810239257349
31 0.210869590848067
32 0.820231266580573
33 0.0933391241255299
34 0.121597597108252
35 0.199447335488201
36 0.114389789751773
37 0.226195197800707
38 0.0737727772509109
39 0.0632202930811867
40 0.0275842619387462
41 0.0879363026294337
42 0.145428339415038
43 0.0675492271631789
44 0.074933428298642
45 0.095016843716855
46 0.16063918629902
47 0.0816128878435569
48 0.100681164875603
49 0.132311418850843
50 0.119052922202076
};
\addlegendentry{$m = 5$}
\addplot [semithick, sienna1408675]
table {%
25 0.0288609620020964
26 0.0318093908089078
27 0.0350804196672376
28 0.0412809854334359
29 0.0521338876414857
30 0.0721134827242666
31 0.108034737472614
32 0.652159228908732
33 0.28785443282468
34 0.180817881826577
35 0.119642993948572
36 0.0817863855972455
37 0.194954973000318
38 0.11500927886744
39 0.110969386457939
40 0.0240701940736222
41 0.0772202248810676
42 0.034418963063546
43 0.0728533044880261
44 0.0190041028071688
45 0.0773714781230674
46 0.0844447991895438
47 0.0532182762865159
48 0.0658708564007379
49 0.0262333507531327
50 0.0211629739354642
};
\addlegendentry{$m = 6$}
\addplot [semithick, orchid227119194]
table {%
25 0.0109115837078727
26 0.0115873352611837
27 0.0123777159016764
28 0.0142136541441188
29 0.0172663360267093
30 0.0253051926074131
31 0.0672034923993614
32 0.879158253171662
33 0.063310055401445
34 0.0646793372150163
35 0.130517395681117
36 0.0892088753164753
37 0.221342491882811
38 0.120421074369457
39 0.113290282457169
40 0.0654657656824433
41 0.0420826135358304
42 0.0720121145563643
43 0.15056505234861
44 0.030460183202324
45 0.0775980883124245
46 0.105440129024401
47 0.0325332986850759
48 0.0405479933051584
49 0.0619230390227052
50 0.00582212109599491
};
\addlegendentry{$m = 7$}
\addplot [semithick, gray127]
table {%
25 0.100979163367256
26 0.102155655335393
27 0.104750713797685
28 0.110394600573759
29 0.122810386732544
30 0.152337800260092
31 0.241012440661112
32 0.703973979247486
33 0.508858162858841
34 0.262392387641122
35 0.23161277714206
36 0.205944117717973
37 0.108622520499135
38 0.149608694635719
39 0.0842306671490624
40 0.00923319231656048
41 0.157905804234083
42 0.0697036105377897
43 0.0971035583891112
44 0.162519587323432
45 0.145739225077711
46 0.183274917495376
47 0.135268439503031
48 0.0937835170639021
49 0.129005773971586
50 0.148163746375044
};
\addlegendentry{$m = 8$}

\draw [thick, dotted] (axis cs: 27, 2.65) -- (axis cs: 27, -0.15) node[pos = 0.65, anchor = center, fill = white, opacity = 0.7, text opacity = 1.0, inner sep = 1pt, red!80!black, fill = white] {$\tau_\mathrm{min}$};
\draw [thick, dotted] (axis cs: 40, 2.65) -- (axis cs: 40, -0.15) node[pos = 0.65, anchor = center, fill = white, opacity = 0.7, text opacity = 1.0, inner sep = 1pt, red!80!black, fill = white] {$\tau_\mathrm{max}$};
\end{axis}

\end{tikzpicture}